\documentclass[amsmath,nobibnotes,aps,superscriptadress,amssymb,pra,showpacs,superscriptaddress,twocolumn,floatfix,nofootinbib,longbibliography,prd,jabr]{revtex4-2}
\usepackage[utf8]{inputenc}
\usepackage[english]{babel}
\usepackage[T1]{fontenc}
\usepackage{tikz}
\usepackage{lipsum}
\usepackage{amsmath}
\usepackage{amsfonts}
\usepackage{graphicx}
\usepackage{multirow}
\usepackage{amssymb}
\usepackage[colorlinks=true,citecolor=blue,linkcolor=blue,urlcolor=blue]{hyperref}
\usepackage{float}
\usepackage{braket}
\usepackage{qcircuit}
\usepackage{multirow}
\usepackage{enumitem} 
\usepackage{bm} 
\usepackage{cancel}
\usepackage{pifont}
\usepackage{anyfontsize}
\usepackage{soul}
\usepackage[left=3cm,right=3cm,top=3cm,bottom=3cm]{geometry}
\usepackage{natbib}
\usepackage{xcolor}

\newcommand{\QFT}{\ensuremath{\hat{\mathcal{F}}}}
\newcommand{\iQFT}{\ensuremath{\hat{\mathcal{F}}^{-1}}}
\newcommand{\FT}{\ensuremath{\mathcal{F}}}

\newcommand{\argmin}{\ensuremath{\text{argmin}}}

\usepackage{amsmath,amsfonts,amssymb,amsthm,graphics,graphicx,epsfig,bbm}
\usepackage{graphicx}
\usepackage{subfigure}
\usepackage{amsmath}
\usepackage{epsfig}
\usepackage{dcolumn}
\usepackage{bm}
\usepackage{color}
\usepackage{times}
\usepackage{epstopdf}
\usepackage{amssymb}
\usepackage{amstext}
\usepackage{latexsym}
\usepackage{hyperref}
\usepackage{amsfonts}
\usepackage{psfrag}
\usepackage{soul,xcolor}
\usepackage[normalem]{ulem}
\usepackage{dsfont}
\usepackage{physics}
\usepackage{footnote}
\usepackage{tikz}
\usepackage{blindtext}
\usepackage{bbm}
\usepackage{url}
\usepackage{textcase}

\begin{document}

\title{Quantum Fourier analysis for multivariate functions and applications to a class of Schrödinger-type partial differential equations}

\author{Paula García-Molina}
\email{paula.garcia@iff.csic.es}
\affiliation{Instituto de Física Fundamental, IFF-CSIC, Calle Serrano 113b, 28006 Madrid, Spain}
\author{Javier Rodríguez-Mediavilla}
\affiliation{Instituto de Física Fundamental, IFF-CSIC, Calle Serrano 113b, 28006 Madrid, Spain}
\author{Juan José García-Ripoll}
\affiliation{Instituto de Física Fundamental, IFF-CSIC, Calle Serrano 113b, 28006 Madrid, Spain}

\begin{abstract}
In this work, we develop a highly efficient representation of functions and differential operators based on Fourier analysis. Using this representation, we create a variational hybrid quantum algorithm to solve static, Schr\"odinger-type, Hamiltonian partial differential equations (PDEs), using space-efficient variational circuits, including the symmetries of the problem, and global and gradient-based optimizers. We use this algorithm to benchmark the performance of the representation techniques by means of the computation of the ground state in three PDEs, i.e., the one-dimensional quantum  harmonic oscillator, and the transmon and flux qubits, studying how they would perform in ideal and near-term quantum computers. With the Fourier methods developed here, we obtain low infidelities of order $10^{-4}-10^{-5}$ using only three to four qubits, demonstrating the high compression of information in a quantum computer. Practical fidelities are limited by the noise and the errors of the evaluation of the cost function in real computers, but they can also be improved through error mitigation techniques.

\end{abstract}

\maketitle

\section{Introduction}

Early at the beginning of quantum computing, one of the suggested applications was the encoding and manipulation of discretized functions\ \citep{Zalka1998,GroverRudolph2002}, as states of the quantum register. This idea opens the field of quantum numerical analysis, where quantum computers assist in tasks such as the solution of linear, nonlinear and differential equations. Kacewicz demonstrated a quantum speed up in the solution of initial-value problems for ordinary differential equations (ODEs)\ \cite{Kacewicz2006}, using quantum amplitude estimation as a subroutine of a classical method. Leyton and Osborne\ \cite{Leyton2008} suggested using quantum computers to solve nonlinear ordinary differential equations. Unlike an earlier work by Kacewicz\ \cite{Kacewicz2006}, later adapted to the Navier-Stokes partial differential equations (PDEs)\ \cite{Gaitan2020}, the quantum computer is not a subroutine in a classical method, but the whole problem is encoded in the quantum computer. Berry\ \cite{Berry2014} pushed this idea forward, transporting the quantum speedups of the Harrow-Hassidim-Lloyd (HHL) algorithm\ \cite{HHL2009} for linear systems of equations, to the solution of ODEs, by means of the Euler method and quantum simulation. Since then, a great deal of effort has been put into this technique, improving its precision\ \cite{Berry2017} and extending the method to PDEs\ \cite{Childs2020}. This technique also suits the finite-element method\ \cite{Montanaro2016} and spectral methods\ \cite{Childs2020b}, as well as a variety of linear problems: the Poisson equation\ \cite{Cao2013}, the heat equation\ \cite{Linden2020} or the wave equation\ \cite{Costa2019}, which is successfully simulated in Ref.\ \cite{Suau2021}. Such ideas, in combination with the Carleman\ \cite{Liu2020} or the quantum nonlinear Schr\"odinger linearization\ \cite{Lloyd2020}, can also be applied to weakly nonlinear differential equations. Note also alternative methods based on hardware-efficient Taylor expansions\ \cite{Xin2020} or on the intrinsic dynamics provided by continuous-variable quantum computers\ \cite{Arrazola2019}.

Even though HHL-based or quantum-simulation-based differential equation solvers exhibit potential quantum speedups, they require large numbers of qubits and operations, far from current noisy intermediate-scale quantum (NISQ) \cite{Preskill2018} devices and closer to the specifications of fault-tolerant scalable quantum computers. In this context, variational hybrid quantum-classical algorithms\ \cite{McClean2016} appeared as a family of methods with lower hardware requirements and strong resilience to noise\ \cite{Sharma2020}, more adequate for the state of the art. In this paradigm, a variational quantum circuit encodes the solution to a complex problem (an ODE or PDE in our case) and the parameters of the circuit are tuned through a learning process that optimizes a loss function.  In some cases, the variational form encodes the complete function\ \cite{Lubasch2020,McArdle2019}, while in others the variational circuit acts as a quantum neural network that, given the right coordinates, outputs a prediction for the function under study\ \cite{Kyriienko2020,Knudsen2020}. The resulting algorithms have a wide range of applications,  including physics\ \cite{Liu2020b,Mocz2021}, chemistry\ \cite{Zhang2020}, and finance \cite{Fontanela2021,Chakrabarti2021,Radha2021}, and may work with both nonunitary\ \cite{McArdle2019,Fontanela2021} and nonlinear differential equations\ \cite{Lubasch2020}.

Both fault-tolerant and NISQ algorithms for quantum numerical analysis need an efficient quantum representation of functions and differential operators to benefit from the quantum paradigm. In this work we propose a different Fourier encoding to map functions and differential operators to the states of an $n$-qubit quantum register, whose representation can be extended with a quantum Fourier interpolation algorithm~\cite{GarciaRipoll2021}. Combined, both tools provide an efficient representation of PDEs and their eigenstates as quantum operators and quantum register states, with errors that can decrease doubly exponentially $O(\exp(-r 2^n))$ in the number of qubits $n$. This favorable scaling motivates the application of these ideas on NISQ hardware, transforming the PDE into a variational principle that can be optimized using both existing and novel variational \textit{Ans\"atze}, that take into account a problem's symmetries.

As a benchmark for these techniques, we consider PDEs of the form
\begin{equation}
  \label{eq:pde}
  \left[D(-i\nabla) + V(\mathbf{x})\right]f(\mathbf{x}) = E f(\mathbf{x}),
\end{equation}
defined over a regular domain $x_i \in [a_i,b_i),$ with periodic boundary conditions, $f(\mathbf{x}+(b_i-a_i)\mathbf{e}_i)=f(\mathbf{x})$ and real functions $D(\mathbf{p}), V(\mathbf{x})\in\mathbb{R}.$ We assume that the PDE is a lower-bounded Hamiltonian operator
\begin{equation}
  \label{eq:Hamiltonian}
  H = D(-i\nabla) + V(\mathbf{x}) \geqslant	 E_{\min}
\end{equation}
and we seek the ground state $E_{\min}$ or lowest-energy excitations using a variational quantum algorithm suitable for solving static PDEs with a Hamiltonian nature. Our study focuses on the impact of finite precision and gate errors in the estimation of the cost function and how this affects the estimation of the solution and the eigenvalues themselves. Surprisingly, we find that the variational algorithm exhibits a great performance, achieving an error $1-F^\infty=10^{-4}-10^{-5}$ in the solution of the harmonic oscillator, and the transmon and flux qubit equations. This precision is well above what one would expect when the cost function is estimated with a finite number of measurements and illustrates the resilience of variational methods when combined with stochastic optimization. Our study shows that to exploit the degree of accuracy and compression offered by our methods requires either high quality qubits and operations or error mitigation techniques. In present quantum computers this is not yet possible and we require a large number of measurements $\sim 10^{4}-10^{5}$ for moderate precision.

The structure of this paper is as follows. In Sec. \ref{sec: Fourier analysis} we propose a set of quantum Fourier analysis tools to efficiently represent functions and operators in a quantum register, as well as an encoding of functions in the form of suitable variational quantum circuits. We provide three variational \textit{Ansätze}: a generic one based on $\sigma^y$ rotations, one adapted for this function encoding\ \cite{Zalka1998,GroverRudolph2002}, and a meta variational circuit that symmetrizes either of those. In Sec. \ref{sec:solver} we propose a variational quantum algorithm to solve Hamiltonian PDEs in order to benchmark the accuracy of these techniques. In Sec.\ \ref{sec:benchmark} we introduce three equations---the quantum harmonic oscillator, the transmon, and the flux qubit---as three models that we will use to run this algorithm. Section\ \ref{sec:numerics} discusses the application of these algorithms for the harmonic oscillator (cf. Sec.\ \ref{sec:numerics-ho}), for the transmon qubit (cf. Sec.\ \ref{sec:numerics-transmon}) and for the flux qubit (cf. Sec.\ \ref{sec:numerics-flux_qubit}). Our study begins with idealized quantum computers, analyzing the limitations of the algorithm, the \textit{Ansatz} and the optimization methods. We then simulate realistic NISQ devices in Sec.\ \ref{sec:numerics-errors}, analyzing the attainable fidelities and introducing error mitigation techniques to improve our estimate of the PDE's eigenvalues. In Sec.\ \ref{sec:conclusion} we summarize, discuss the conclusions drawn from this work, and outline further research. The performance of the algorithm is still subpar with other classical methods for existing hardware. However, it serves its purpose to verify the high precision of the quantum Fourier analysis methods, even for a low number of qubits (three to four qubits).

\section{Quantum Fourier analysis for multivariate functions}
\label{sec: Fourier analysis}
\subsection{Position space discretization}

In quantum numerical analysis it is key to find an efficient representation of functions in a quantum register. Without loss of generality, we center our discussion on one-dimensional problems. Our work assumes periodic functions or functions $f(x)$ that vanish towards the boundaries of a finite interval  $[a,b),$ of size $L_x=|b-a|$. We also focus on functions that are bandwidth limited: i.e., their Fourier transform $\tilde{f}(p)=[\FT{f}](p)$ is negligible outside a corresponding interval in momentum or frequency space $[-L_p/2,L_p/2).$

With each such function $f(x)$ we associate a quantum state $\ket{f^{(n)}}$ with $n$ qubits, discretizing the function on a regular grid with $2^n$ points, labeled
\begin{equation}
  \label{eq:position-grid}
  x_s^{(n)} = a + s \Delta{x},
\end{equation}
with $s\in\{0,1,\ldots,2^n-1\}$ and $\Delta{x}^{(n)} = \frac{L_x}{2^n}$. The discretized and normalized state is a linear superposition
\begin{equation}
  \ket{f^{(n)}} = \frac{1}{\mathcal{N}_f^{1/2}} \sum_{s=0}^{2^n-1} f(x_s^{(n)})\ket{s}
\end{equation}
of quantum register states $\ket{s}$ that encode the integer $s$ into the states of $n$ qubits, with a normalization constant $\mathcal{N}_f$ that depends on the number of qubits. It is important to remark that, as found in Refs.\ \cite{Zalka1998,GroverRudolph2002}, in order to describe a function with a fine grid we only need a logarithmically small number of qubits $n=O(\log_2(1/\Delta{x})).$ This advantage translates to the scaling of approximation errors in the overall method, as discussed in Sec.\ \ref{sec:errors}.

The Nyquist-Shannon theorem \cite{Nyquist1928,Shannon1949} ensures that any bandwidth-limited function $f(x)$ can be interpolated from a discretization $\ket{f^{(n)}}$ with spacing $\Delta{x}^{(n)}\leqslant{2\pi}/L_p,$ up to exponentially small errors. This allows us to (i) estimate the smallest number of qubits required to make the sampling accurate, (ii) establish the inverse mapping, from states $\ket{f^{(n)}}$ to functions,  and (iii) develop an algorithm to create interpolated states $\ket{f^{(n+m)}}$ and $\ket{\tilde{f}^{(n+m)}}$ for the estimation of the wavefunction and its energy.

\subsection{Momentum space and Fourier interpolation} \label{sec: Fourier interpolation}

Given an $n$-qubit discretized function $\ket{f^{(n)}}$, we can straightforwardly implement the quantum Fourier transform (QFT). This unitary operator $\QFT$ is the quantum analog of the discrete Fourier transform,
\begin{equation}
  \ket{r}\mapsto\frac{1}{\sqrt{2^n}}\sum_{s=0}^{2^n-1}e^{i 2\pi r s/2^n}\ket{s}.
\end{equation}
When applied on $\ket{f^{(n)}},$ $\QFT$ produces the quantum state that encodes the discrete Fourier transform of the series of values $\{f(x_s)\}$
\begin{align}
  \ket{\tilde{f}^{(n)}} &= \sum_s \tilde{f}^{(n)}(p_s)\ket{s} := \QFT\ket{f^{(n)}} \\
  &= \frac{1}{\sqrt{2^n}}\sum_{r,s=0}^{2^n-1}e^{i 2\pi s r/2^n}f(x_r)\ket{s}.\notag
\end{align}
As in the discrete Fourier transform, we have to deal with the annoying ordering of quasimomenta $p_s\in[-L_p/2,L_p/2)$ that stores negative frequencies in the higher states of the quantum register
\begin{equation}
  \label{eq:momenta}
  p_s = \frac{2\pi}{\Delta{x}^{(n)}2^n} \times \left\{
    \begin{array}{ll}
      s &  \text{for } 0\leqslant s < 2^{n-1}\\
      s-2^n & \mbox{otherwise.}
    \end{array}
  \right.
\end{equation}

If we work with bandwidth-limited functions and the discretization step is small enough $\Delta{x}^{(n)}\leqslant 2\pi/L_p,$ we can reconstruct the original function from the discretized state in momentum space. Up to normalization, we write
\begin{equation}
  \label{eq:continuous-position}
  f(x) \propto \sum_{s=0}^{2^n-1} e^{-ip_s x} \langle{s|\QFT |f^{(n)}}\rangle.
\end{equation}
This equation represents a mapping from states in the quantum register to continuous, infinitely differentiable bandwidth-limited functions. It also provides us with a recipe to interpolate the discretized function $\ket{f^{(n)}}$ up to arbitrary precision, using more qubits to represent more points in the position space.

\begin{figure}[t]
\includegraphics[width=0.45\textwidth]{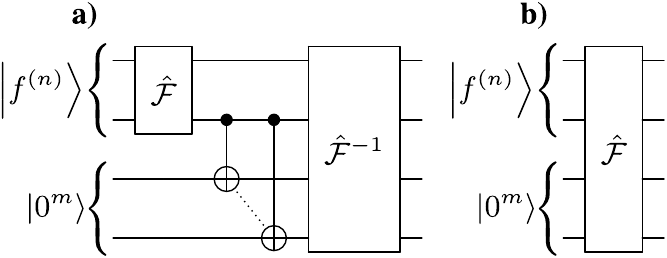}
\caption{Quantum interpolation algorithms. (a)  Algorithm to recreate a finer interpolation in position space $\ket{f^{(n+m)}}$ adding $m$ qubits to a previous discretization $\ket{f^{(n)}}.$  (b) Algorithm to recreate a momentum space discretization with $n+m$ qubits from $\ket{f^{(n)}}.$}
  \label{fig:interpolation}
\end{figure}

As sketched in Fig.\ \ref{fig:interpolation}(a), to perform this position space interpolation scheme we need three steps. First, compute the QFT of the originally encoded function $\ket{\tilde{f}^{(n)}}.$ Second, add $m$ auxiliary qubits to enlarge the momentum space. Due to the anomalous encoding of momenta [Eq.\ \eqref{eq:momenta}], the original discretization with $2^n$ points must be mapped to the intervals $s\in [0,2^{n-1})\cup[2^{n+m}-2^{n-1},2^{n+m})$. This is done using the operation $U_\text{sym}$ where the sign of the values of the auxiliary register is determined by the most significative qubit of the original register. Finally, we Fourier transform back to recover the state with $n+m$ qubits. The complete algorithm reads
\begin{align}
  \label{eq:interpolation}
  |f^{(n+m)}\rangle &= \iQFT{}U_{\text{sym}} \left(\ket{0}^{\otimes m} \otimes \QFT\ket{f^{(n)}}\right)\\
  &=: U_\text{int}^{n,m}\ket{f^{(n)}}.\notag
\end{align}
The interpolation algorithm reveals the exponential advantage provided by the encoding of the function in the quantum register. A similar interpolation executed in a classical computer would require $O(2^{n+m})$ real values, processed with a fast Fourier transform that demands $O((n+m) 2^{n+m})$ operations. In contrast, the quantum computer uses just $n+m$ qubits and $O((n+m)^2)$ quantum gates.

We can define another, simpler momentum space interpolation scheme that associates $\ket{f^{(n)}}$ with a continuous differentiable function in momentum space $\tilde{f}(p)$,
\begin{align}
  \label{eq:continuous-momentum}
  \tilde{f}(p) &\propto \sum_{s=0}^{2^n-1} e^{ip x_s} \langle{s|f^{(n)}}\rangle,\;p\in[-L_p/2,L_p/2).
\end{align}
The quantum interpolation method enlarges the number of points in momentum space by a factor $2^m$ using $m$ auxiliary qubits and a quantum Fourier transform [Fig.\ \ref{fig:interpolation}b]
\begin{equation}
  \ket{\tilde{f}^{(n+m)}} = \QFT_{n+m}\left[\ket{0}^{\otimes m} \otimes \ket{f^{(n)}}\right].
\end{equation}
This is equivalent to extending the grid on which we discretize $f(x)$ from $[a,b)$ to $[a,a+2^mL_x),$ setting all those extra points to the same value that the function takes at the boundary. This extension increases the interval size while preserving the spacing, which causes the grid in momentum space to become denser,  $\Delta{p}\to \Delta{p}/2^m,$ over the same frequency domain $[-L_p/2,L_p/2).$

As a corollary, the study of the two-way interpolation schemes and the Nyquist-Shannon theorem provides us with the optimal sampling or discretization for a function, given the domain sizes $L_p$ and $L_x$ in momentum and position space, respectively. Our argument is that for the sampling theorem to succeed and to provide us with good interpolations~\eqref{eq:continuous-position} and \eqref{eq:continuous-momentum}, the spacings in momentum and position need to satisfy $\Delta{x}^{(n)}\leqslant 2\pi/L_p$ and $\Delta{p}^{(n)}\leqslant 2\pi/L_x.$ This is achieved with a minimum number of qubits, given by
\begin{equation}
  \label{eq:nqubits}
  n_{\min}\leqslant \log_2\left(\frac{L_pL_x}{2\pi}\right).
\end{equation}
We use this estimate in our numerical studies and following sections.

\subsection{Approximation errors} \label{sec:errors}

As discussed in the Appendix A, the quantum states that we construct are interpolants of the solution $f^{(n)} = I_N f,$ with $N=2^n$ the number of interpolation points. The error made by such interpolation originates in (i) the truncation of the Fourier series to $N$ modes and (ii) the computation of this Fourier series using discrete Fourier transforms. When the interpolated function is differentiable up to order $m$ (and in this case we know $m\geqslant	 2$), the error of the method is expected to be $O(N^{-m})\sim O(2^{-nm}),$ which is close to other quantum algorithms using finite-difference approximations of order $m$.

When the solution is almost everywhere differentiable up to order $m$ but the function experiences discontinuities, we may find additional errors due to a phenomenon called Gibbs oscillation. As discussed in the Appendix A, those errors are localized around the discontinuity, and affect our estimation of $f(x)$, but because they have zero measure as $N\to\infty$, the errors we make when estimating the energy and various observables decay in a favorable way with the refinement of the discretization.

In practice, in many physically motivated PDEs, including the ones we discuss below, we find that solutions are analytical functions and therefore arbitrarily differentiable. In this case, the Fourier method provides an approximation that is exponentially good in the size of the lattice, and doubly exponential in the number of qubits $O(e^{-rN})\sim O(e^{-r 2^n}),$ with some problem-dependent constant $r$. Moreover, we only find Gibbs oscillations when we approximate problems that are defined over the whole real line. As we will see in Sec.\ \ref{subsec: ho}, those errors can be made arbitrarily small by enlarging the interval size $L$ until the value of $f(x)$ is negligible, making the function and higher derivatives almost periodic on the computation interval.

\subsection{Dimensionality}

The study for one dimension can be generalized to multidimensional problems with similar tools and similar considerations for the precision of the method. To represent a $d$-dimensional function $f(\boldsymbol{x}) = f(x_0, x_1,\dots, x_{d-1})$, we encode each variable in an $n_i$-qubit register by discretizing them on a regular grid with $2^{n_i}$ points, labeled
\begin{equation}
  \label{eq:multi-position-grid}
  x_{i,{s_i}}^{(n_i)} = a + s_i \Delta{x}_i,
\end{equation}
with $s_i\in\{0,1,\ldots,2^{n_i}-1\}$, where $i = 0,\dots,d-1$ labels the dimension of the coordinate and $s_i$ is an integer encoded in $n_i$ qubits. This allows us to store the $d$-dimensional function $f(\boldsymbol{x})$ in an $n=\sum_i n_i$ qubit quantum register as
\begin{equation}
  \ket{f^{(n)}} = \frac{1}{\mathcal{N}_f^{1/2}} \sum_{\lbrace s_i\rbrace} f(\boldsymbol{x})\ket{\boldsymbol{s}},
\end{equation}
where $\boldsymbol{s} = (s_0,s_1,\dots, s_{d-1})$ is the collection of $d$ integers encoded in the quantum register.

To represent a function in momentum space we resort to the multidimensional QFT
\begin{equation}
  \ket{\boldsymbol{r}}\mapsto\frac{1}{\sqrt{2^n}}\sum_{\lbrace s_i \rbrace}e^{i 2\pi r \sum_i s_i/2^n}\ket{\boldsymbol{s}},
\end{equation}
which extends the one-dimensional QFT to the $d$ dimensions of $f(\boldsymbol{x})$ by applying the corresponding quantum circuits on the $n_i$ qubits encoding each dimension $x_i$. We can similarly apply this reasoning to extend the quantum Fourier interpolation to higher dimensions. Moreover, in order to compute a partial derivative $\partial_{x_i}$ it is only necessary to apply the QFT on the $n_i$ qubits of the register, and hence the cost of the QFT and its inverse is the same as for the one-dimensional case.

\subsection{Variational quantum circuits} \label{sec: ansatz}

In the previous sections we have discussed how to code and manipulate functions in a quantum register using quantum Fourier analysis. In practical applications we still need to construct those functions in a quantum register as a combination of gates. As discussed in the literature, a possible route for this is to engineer generic variational circuits which can be specialized to each problem. In the following we review an \textit{Ansatz} commonly used in the variational quantum eigensolver (VQE) literature\ \cite{Kandala2017}, we introduce a variational circuit inspired by an exact representation of functions\ \cite{Zalka1998,GroverRudolph2002}, and we introduce one technique to incorporate symmetries into both of these \textit{Ansätze}.

\subsubsection{$R_Y$ \textit{Ansatz}}
As a baseline for our study, we use a variational \textit{Ansatz} that combines controlled-NOT (CNOT) gates with local, real-valued transformations on the qubits generated by the $\sigma^y$ operator [c.f. Fig.\ \ref{fig:ansatz}(a)]. The result is a parametrized unitary
\begin{align} \label{eq: RY}
  W({\bm\theta}) &= \prod_{q=0}^{n-1} R^y_q\left(\theta_q^{\text{depth}+1}\right) \\
    & \times \prod_{d=1}^{\text{depth}}\left[\prod_{c=0}^{n-1}\prod_{c<t} \text{CNOT}_{c, t} \prod_{q=0}^{n-1} R^y_q\left(\theta_q^d\right)\right], \notag
\end{align}
where the parameters are the angles of the $R_Y(\theta)=\exp(-i\theta\sigma^y/2)$ rotations. This variational \textit{Ansatz} is available in many quantum computing frameworks, including QISKIT\ \cite{Qiskit}, which is the one used for our simulations.

\subsubsection{Zalka-Grover-Rudolph (ZGR) \textit{Ansatz} }

We can derive a better variational \textit{Ansatz} for real functions using the ideas from Zalka\ \cite{Zalka1998} and Grover and Rudolph\ \cite{GroverRudolph2002} to discretize non-negative probability distributions in a quantum register. In their work they showed that probability distributions could be approximated by conditional rotations of the least significant qubits based on the state of all previous qubits,
\begin{equation}
  \ket{f^{(n)}_{\bm\theta}} = \prod_{i=n-1}^1 \prod_{z_i=0}^{2^{i-1}-1} \exp[i\theta_{i}(z_i)\sigma^y_i\ket{z}\!\bra{z})]\ket{0}^{\otimes n}.
\end{equation}
Here, $z_i$ is an integer constructed from the $i$ first qubits $z=s_0s_1\cdots s_{i-1}$ and $\theta_i(z_i)$ is a collection of angles designed to reproduce the desired function. Note how this is a constructive process where the state of a qubit with low significance $i$ is determined by controlled rotations defined by the more significant qubits $j \in\{0,1,\ldots,i-1\}.$

\begin{figure}[t]
  \includegraphics[width=0.45\textwidth]{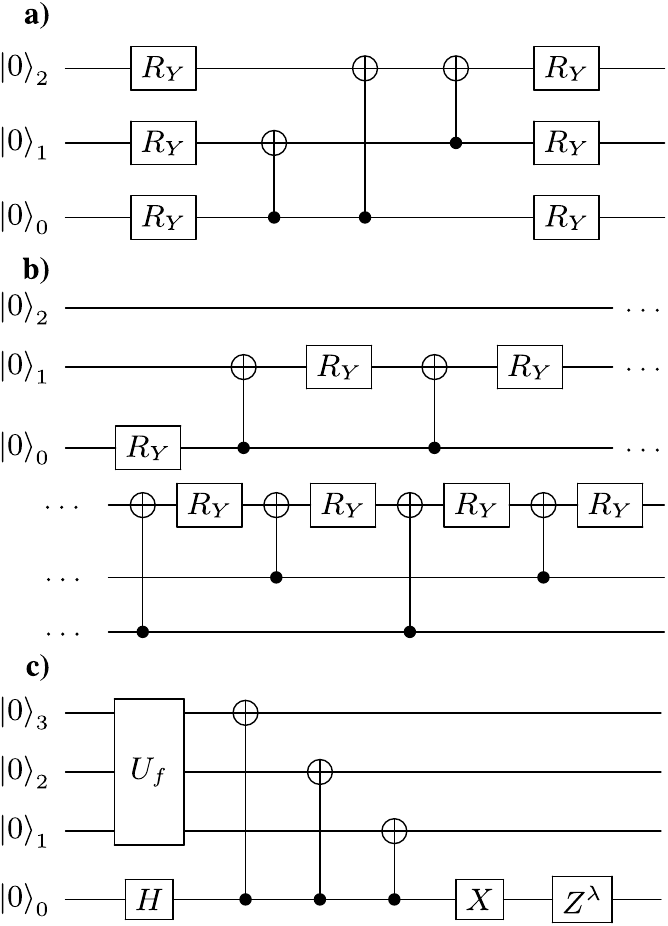}
  \caption{Variational representation of one-dimensional smooth functions. (a) $R_Y$ \textit{Ansatz} of depth one, with full entanglement over three qubits. (b) ZGR \textit{Ansatz} to represent a function $f(x)$ with three qubits. (c) Ansatz to represent an (anti)symmetric function $g(x) = (\text{sgn}\,{x})^{\text{sym}}f(|x|),$ where $U_f$ encodes $f(x>0)$.}
  \label{fig:ansatz}
\end{figure}

In practice, the construct from Zalka-Grover-Rudolph (ZGR) could be implemented using $\sigma^y$ rotations and CNOT gates. Inspired by this idea, we designed an alternative variational \textit{Ansatz} [c.f. Fig.\ \ref{fig:ansatz}b] where such rotations are parametrized by a similar number of angles and is slightly more efficient
\begin{equation}\label{eq: ZGR}
  \ket{f^{(n)}_{\bm\theta}} = \prod_{i=n-1}^{1} \prod_{z_i=0}^{2^{i-1}-1} e^{i\theta_{iz_i}\sigma^y} N(z,i) e^{i\theta_{00}\sigma^y_0}\ket{0}^{\otimes n}.
\end{equation}
The $N$ is a CNOT rotation of the $i$th qubit, controlled by a $j$ qubit which is determined by the most significant bit that is active in $z\mbox{ XOR }(z-1).$ As in the previous construct, the exponentially growing number of parameters makes this an ill-advised choice for very large computations, but we use it because (i) it illustrates the required ordering of conditional operations, from high to low-significant bits, (ii) it is more accurate than the pure $R_Y$ \textit{Ansatz}, and (iii) in practice it gets extremely close to the exact construct, but does not require additional NOT gates.

\subsubsection{Symmetrization of variational circuits}

We can significantly reduce the number of variational parameters by embedding the symmetries of the function in the circuit. Let us consider discretized functions with reflection symmetry, $f(x)=(-1)^\lambda{f}(-x)$, with $\lambda=0,1.$ This symmetry can be added to the variational \textit{Ansatz}, engineering first a unitary $U_f$ that describes the function in the positive sector $0\leqslant x \leqslant L_x/2$ with $n-1$ qubits and applying unitary operations that extend this state to $n$ qubits, covering also the negative part of the discretization $-L_x/2\leqslant x \leqslant L_x/2.$

For this encoding to succeed, we first need to impose a discretization of the space that is also symmetric, which we do by slightly changing the relation between qubit states and positions
\begin{equation}
  \label{eq:symmetric-x}
  x_s^{(n)} = -\frac{L_x}{2}+\left(s+\frac{1}{2}\right)\Delta{x}.
\end{equation}
With this, our symmetrized function satisfies
\begin{equation}
  \langle{1s_1\dots s_{n-1}|f^{\lambda}}\rangle =
  (-1)^\lambda\langle{0\bar{s}_1\dots \bar{s}_{n-1}|f^\lambda}\rangle.
  \label{eq:symmetrization}
\end{equation}
Note how positive coordinates $x_s>0$ relate to the negative ones $x_s<0$. These relations can be embedded into the variational circuit as shown in Fig.\ \ref{fig:ansatz}(c). First, we create the most significant bit of the wavefunction $s_n$ on a quantum superposition of both $s_0=1$ ($x_s>0$) and $s_0=0$ ($x_s<0$). We then create the encoded function for the positive valued coordinates in the $n-1$ least significant qubits. Finally, we reverse the orientation of the function in the negative coordinates~\eqref{eq:symmetrization} and set the right sign for the encoded state.

Other symmetrization methods can be found in the literature, such as the one in Ref.\ \cite{Seki2020}. This work presents a symmetry-adapted VQE using a projection operator, whose nonunitarity leads to a classical postprocessing. Our approach symmetrizes the state by adding one qubit to the circuit and a small number of single- and two-qubit gates, which is a fully coherent approach and does not require any postprocessing.

\section{Variational quantum PDE solver}\label{sec:solver}

We will apply the quantum Fourier analysis techniques from Sec.\ \ref{sec: Fourier analysis} to develop a hybrid quantum-classical algorithm for solving PDEs of the form\ \eqref{eq:pde}, which can be rewritten as a lower-bounded Hamiltonian operator\ \eqref{eq:Hamiltonian}. As practical examples of this type of equations, we will study the one-dimensional quantum harmonic oscillator (see Sec.\ \ref{subsec: ho}),  and the transmon (Sec.\ \ref{subsec: transmon}) and flux (Sec.\ \ref{subsec: flux qubit}) qubits.

\begin{figure}[t]
\centering
\includegraphics[width=0.5\textwidth]{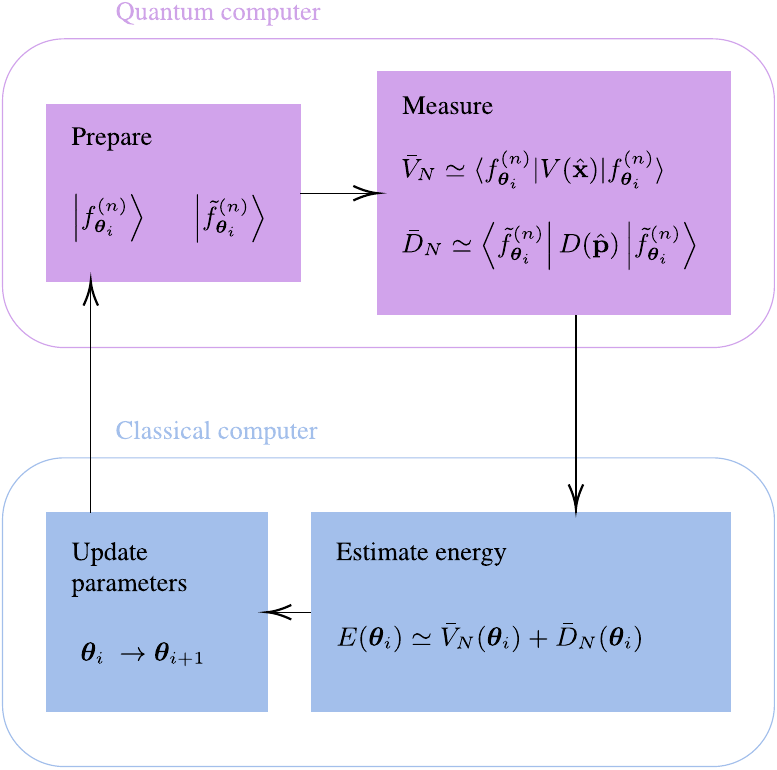}
\caption{Variational quantum PDE solver. We use a quantum computer to initialize two quantum circuits in the states $\ket{f^{( n)}_{\bm{\theta }_{i}}}$ and $\ket{\tilde{f}^{( n)}_{\bm{\theta }_{i}}}.$ We estimate the expectation values of the $V(\hat{\mathbf{x}})$ and $D(\hat{\mathbf{p}})$ operators from measurements in the quantum computer, thereby approximating the energy functional $E[\bm{\theta}]$. A classical computer uses these estimates to iteratively update the parameters of the variational quantum circuit, until convergence.} \label{fig:VQE}
\end{figure}

The ingredients of the algorithm are (Fig.\ \ref{fig:VQE}) (i) a map between states $\ket{f^{(n)}}$ of a quantum register with $n$ qubits and bandwidth-limited continuous functions $f(\mathbf{x})$; (ii) the realization that given $\ket{f^{(n)}},$ the QFT creates a state $\ket{\tilde{f}^{(n)}}$ that encodes the classical Fourier transform $\tilde{f}=\FT{f}$ of the encoded function $f(\mathbf{x})$; (iii) a quantum algorithm that uses $\ket{f^{(n)}}$ and $\ket{\tilde{f}^{(n)}}$ and suitable representations of position and momentum operators $\hat{\mathbf{x}}$ and $\hat{\mathbf{p}}$, to estimate the energy functional $E[f]=(f,Hf)$ with polynomial resources as
        \begin{align} \label{eq:E}
            E[f] &= \langle{\tilde{f}^{(n)} | D(\hat{\mathbf{p}}) | \tilde{f}^{(n)}}\rangle \\
            &+ \langle{f^{(n)} | V(\hat{\mathbf{x}}) | f^{(n)}}\rangle;\notag
        \end{align}
(iv) a variational quantum circuit $W(\bm\theta),$ creating parametrized states of a quantum register with $n$ qubits $\ket{f^{(n)}_{\bm\theta}}=W(\bm\theta)\ket{00\ldots 0}$ (Sec. \ref{sec: ansatz}); and (v) a classical optimization algorithm, such as constrainded optimization by linear approximation (COBYLA), simultaneous perturbation stochastic approximation (SPSA), or adaptative moment estimation (Adam), that given $E(\bm\theta)$ finds the parameters $\bm\theta$ that minimize this function. Ideas (i) and (iv) establish a map from a set of real optimizable parameters ${\bm\theta}\in\mathbb{R}^k,$ to the set of bandwidth limited functions $f_{\bm\theta}(\mathbf{x})$ and their energies $E[f_{\bm\theta}]=:E(\bm\theta).$  Algorithms (ii) and (iii) provide us with a quantum recipe to estimate $E(\bm\theta).$ The classical algorithm (v) hybridizes with the quantum algorithms (i)-(iv), allowing us to search for the variational function $f_{\bm\theta}$ that best approximates a solution to Eq.\ \eqref{eq:pde},
\begin{equation}\label{eq: solution}
  \argmin_f \langle{f|H|f}\rangle\leqslant f_{\argmin_{\bm\theta} E(\bm\theta)}.
\end{equation}

Using the association between states in a quantum register and the space of bandwidth-limited differentiable functions in Sec.\ \ref{sec: Fourier analysis}, we can discuss the Hamiltonian PDEs\ \eqref{eq:pde}. We associate with the differential operator\ \eqref{eq:Hamiltonian} a quantum representation $\hat{H}^{(n)} = D(\hat{p}^{(n)}) + V(\hat{x}^{(n)})$ where the position and differential operators  $V(\hat{x})$ and $D(\hat{p})$ are defined by
\begin{align}
  \label{eq:operators}
  V(\hat{x}^{(n)}) &:= \sum_s V(x_s)\ket{s}\!\bra{s},\\
  D(\hat{p}^{(n)}) &:= \iQFT \sum_s D(p_s)\ket{s}\!\bra{s} \QFT.
\end{align}

With this identification, we can translate the search for the stationary solutions~\eqref{eq:pde} to the quest for eigenstates of the $\hat{H}$ operator, employing the VQE\ \cite{Peruzzo2014,McClean2016}. Using a dense variational family of unitary operators $W(\bm{\theta})$ acting on $n$ qubits, we define a continuous family of trial states $\ket{f_{\bm\theta}^{(n)}}:=W(\bm\theta)\ket{0}^{\otimes n}$ and a cost function
\begin{equation}
  \label{eq:Etheta}
  E(\bm\theta) := \langle{f_{\bm\theta}^{(n)}|\hat{H}^{(n)}|f_{\bm\theta}^{(n)}}\rangle.
\end{equation}
In order to approximate the solution to Eq.~\eqref{eq:pde} with minimal eigenvalue  $E_\text{min},$ we will use the continuous function associated with the quantum state $\ket{f_{\bm\theta_\text{min}}^{(n)}}$ that results from the variational search
\begin{equation}
  \bm\theta_\text{min} := \text{argmin}\; E(\bm\theta).
\end{equation}

We implement this search as a hybrid algorithm, with a classical algorithm that optimizes $E(\bm\theta)$ using the estimations of $E(\bm\theta)$ provided by a quantum computer or a simulator thereof. In the NISQ scenario we do not have access to the exact value of $E(\bm\theta)$ but a randomized estimator that results from a finite set $M$ of measurements. In this work we estimate this ``energy'' as the sum of two random variables, one arising from measurements of the position and another one from the momentum operator
\begin{equation}
  E(\bm\theta)\simeq \bar{E}_M  := \bar{V}_M + \bar{D}_M + O\left(\frac{1}{\sqrt{M}}\right).
\end{equation}
Note that the values of $\bar{V}_M$ and $\bar{D}_M$ are computed separately, creating the quantum states $\ket{f_{\bm\theta}^{(n)}}$ and $\ket{\tilde{f}_{\bm\theta}^{(n)}}:=\QFT\ket{f_{\bm\theta}^{(n)}}$ in different experiments and measuring those states in the computational basis. This approach leads to statistical uncertainties that are of the order $\Delta\hat{V}/\sqrt{M}$ and $\Delta\hat{D}/\sqrt{M}.$ Better algorithms could be constructed by using amplitude estimation over approximate implementations of the unitary operator $\exp(-i\hat{H}\Delta{t})$, but this requires an infrastructure and a precision of gates that is not presently available.

Because of the way that we associate continuous, infinitely differentiable functions $f^{(n)}(x)$ with quantum states $\ket{f^{(n)}}$, the formula\ \eqref{eq:Etheta} gives us the exact value of the functional $E[f^{(n)}]=(f^{(n)}, H f^{(n)}).$ This means that our variational algorithm finds a strict upper bound on the exact eigenvalue $E_\text{min}$ and the errors of the method can only be due to the expressive power of the variational \textit{Ansatz} and the capacity of $f^{(n)}(\mathbf{x})$ to approximate the solution $f(\mathbf{x})$.

Finally, note that a related method has been implemented in Ref.\ \cite{Chakrabarti2021} for solving the harmonic oscillator equation and engineering Gaussian states in quantum circuits.

\color{black}
\section{Benchmark equations}
\label{sec:benchmark}

We will benchmark our variational quantum PDE solver using  three important equations: the quantum harmonic oscillator\ \eqref{eq:oscillator} and the equations for the transmon\ \eqref{eq:transmon} and flux\ \eqref{eq: flux qubit} qubits. They produce simple, highly differentiable functions that can be analytically computed for the two first examples. However, they also involve different boundary conditions, which makes the practical study a bit more interesting.

\subsection{Harmonic oscillator} \label{subsec: ho}

The Schrödinger equation of an harmonic oscillator of mass $m$ and angular frequency $\omega$ is
\begin{equation}
  \label{eq:oscillator}
  \left(-\frac{\hbar^2}{2m}\partial_x^2 + \frac{1}{2}m\omega^2 x^2 -E\right)f(x)=0.
\end{equation}
The exact solutions to this problem are given by
\begin{equation}
    f_n(x)=\left(\frac{\beta^2}{\pi}\right)^{1/4}\frac{1}{\sqrt{2^n n!}}e^{-\beta^2x^2/2} H_n(\beta x),
\end{equation}
where $\beta=\sqrt{m\omega/\hbar}$ and $H_n$ is the Hermite polynomial of order $n$. For $m=\omega=\hbar=1$, we obtain the ground state of the harmonic oscillator, which is a trivial Gaussian
\begin{equation}\label{eq:eq_ho}
    f_0(x) = \frac{1}{\pi^{1/4}}e^{-x^2/2}.
\end{equation}
This function is real, symmetric and even, and is thus particularly well suited for the variational \textit{Ans\"atze} discussed above.

In our quantum numerical analysis, we constrain Eq.\ \eqref{eq:oscillator} to a finite domain $\;|x|\leqslant L_x/2,$ defined symmetrically around the origin as in\ \eqref{eq:symmetric-x}. Following the prescriptions from the Nyquist-Shannon theorem, we vary the length of the interval according to the number of available qubits, as $L_x\approx\sqrt{2\pi2^n},$ to maximize the accuracy in both position and momentum space.

\subsection{Transmon qubit} \label{subsec: transmon}
The eigenstates for a superconducting transmon qubit\ \cite{Koch2007} without charge offset are obtained by solving the equation
\begin{equation}
  \label{eq:transmon}
  \left[-4 E_C\partial_\varphi^2 - E_J \cos(\varphi) -E\right]f(\varphi)=0.
\end{equation}
The phase variable is periodic over the interval $\varphi\in[-\pi,\pi)$. The model is parametrized by the Josephson energy $E_J$ and the capacitive energy $E_C,$ which we choose to be $E_C= E_J/50$. The eigenfunctions of the transmon qubit are given by the Mathieu functions, analytical solutions of Mathieu's differential equation
\begin{equation}\label{eq:eq_transmon}
  \frac {d^{2}y}{dx^{2}}+(a-2q\cos 2x)y=0.
\end{equation}

If we expand the problem around $\varphi=0$, in the limit of large $E_J/E_C$,
\begin{equation}
  H= -4E_C \partial_\varphi^2 + \frac{1}{2}E_J \varphi^2,
\end{equation}
which behaves like an harmonic oscillator with
\begin{equation}
  \frac{\hbar^2}{2m}\sim 4E_C, \quad \frac{1}{2}E_J\sim \frac{1}{2}m\omega^2,
\end{equation}
and effective frequency $\hbar\omega=\sqrt{8E_CE_J}$. Therefore, in some regimes the transmon ground state can be approximated by a Gaussian function
\begin{equation}
  \psi(\varphi) \propto \exp\left[-\frac{1}{2}\left(\frac{\varphi}{a_0}\right)^2\right]
\end{equation}
in the dimensionless variable $x=\varphi/a_0$ with unit length
\begin{equation}
  a_0^4 = \frac{8E_C}{E_J}.
\end{equation}
The Gaussian approximation is not perfect, as it fails to capture the nonlinear contributions and does not take into account the periodicity of the function $f(\varphi).$ In particular, unlike the case of the harmonic oscillator, we are not free to choose the interval length, which is fixed to $2\pi.$

\subsection{Three-junction flux qubit}\label{subsec: flux qubit}

As the third model in our study, we consider the one-dimensional reduction of the model for a three-junction flux qubit. This problem is given by a Schrödinger equation for which there is no analytical solution
\begin{equation} \label{eq: flux qubit}
    \left[-\frac{E_c}{\frac{1}{2}+\alpha}\partial_\varphi^2-E_J\left[2\cos(\varphi)-\alpha\cos(2\varphi)\right]-E\right]f(\varphi)=0.
\end{equation}
In this model, the small junction size takes typical values $\alpha \sim 0.7-0.8.$ In this scenario, the inductive potential $V(\varphi)=-E_J(2\cos(\varphi)-\alpha\cos(2\varphi))$ develops two minima. The ground state of the qubit is a wavefunction delocalized between those minima. The function $f(\varphi)$ is symmetric, even, periodic and vanishes to zero towards the boundaries of the interval, which makes it adequate for our method. For the simulations we have chosen an inductive-to-capacitive energy ratio $E_J/E_C = 50$ and a junction size $\alpha = 0.7.$ These values produce a qubit with a gap that is comparatively smaller than that of the transmon qubit discussed before, thus requiring better estimates of the energy functional to distinguish the ground from the excited states.

\color{black}
\section{Numerical results}
\label{sec:numerics}

In this section we study the application of our variational quantum PDE solver to the harmonic oscillator, and the transmon and flux qubits. We use the ZGR and the $R_Y$ variational \textit{Ansätze}, searching for the ground-state solution and energy for two, three, four, five, and six qubits. The first part of this study combines an idealized quantum computer simulated by QISKIT\ \cite{Qiskit},  with three classical optimizers: COBYLA, a gradient-free method; SPSA, a stochastic optimizer with numerical gradient; and Adam, which we combine with an analytic estimate of the gradient\ \cite{Mitarai2018, Schuld2019}. In Sec.\ \ref{sec:numerics-errors} we discuss how the algorithm performs in a more realistic scenario with errors, analyzing how these errors affect both the evaluation of the function and its properties.

\begin{figure}[t]
\centering
\includegraphics[width=0.50\textwidth]{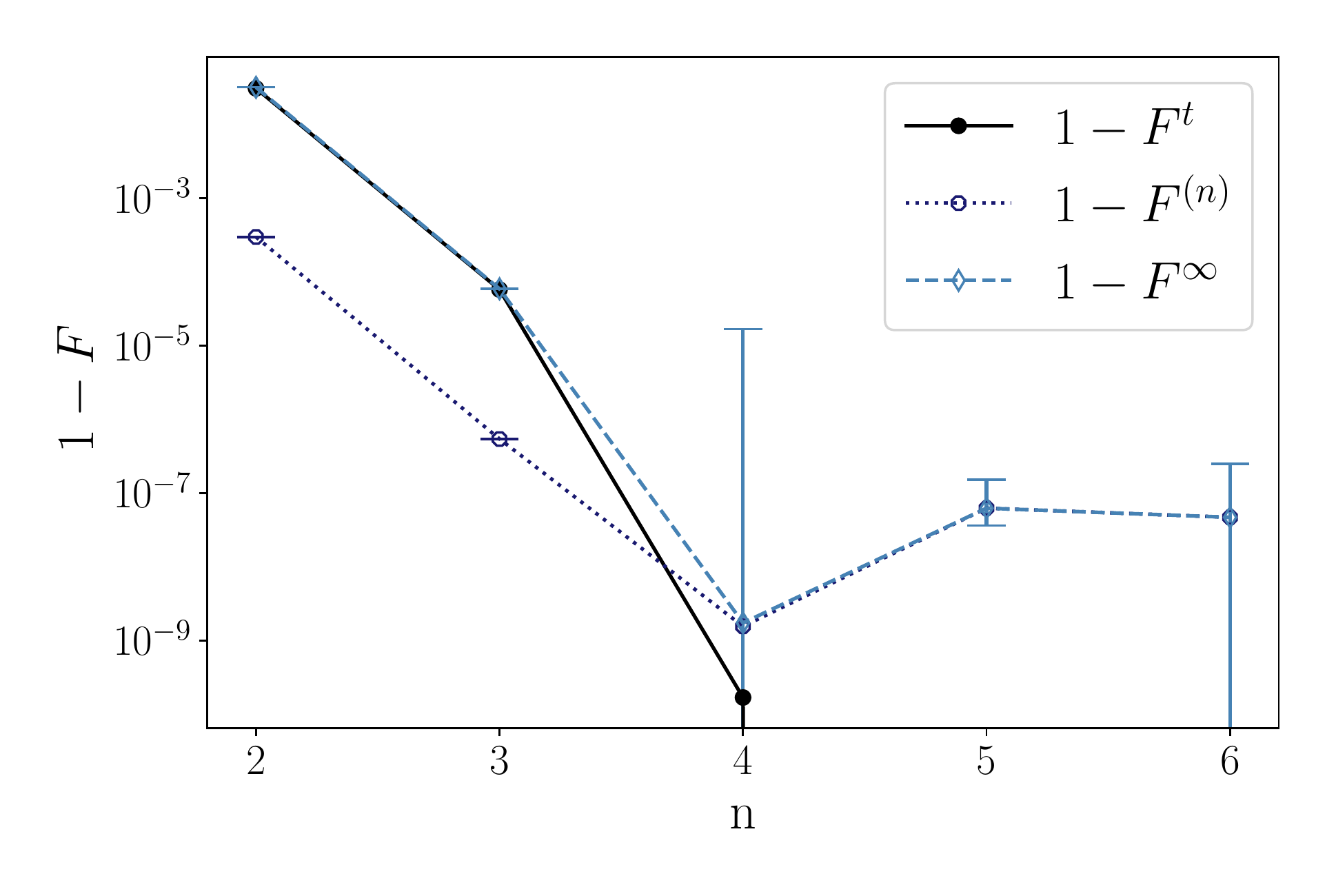}
\caption{Comparison of the infidelity figures of merit (the median and the standard deviation around the mean of 100 repetitions) for the ZGR \textit{Ansatz} for the harmonic oscillator for the numerical limit. The theoretical infidelity $1-F^t$ is the infidelity of the theoretical $2^n$-point wavefunction interpolated up to $2^{12}$ points and the theoretical $2^{12}$-point wavefunction. The $n$-qubit infidelity $1-F^{(n)}$ is the infidelity of the $n$-qubit function obtained from the optimization and the $2^n$-point theoretical function. The continuous infidelity $1-F^\infty$ is the infidelity of the $n$-qubit function interpolated up to $2^{12}$ points and the $2^{12}$-point theoretical function}
\label{Fig:infid_comparison}
\end{figure}

\begin{figure*}[hbt!]
\centering
\includegraphics[width=1\textwidth]{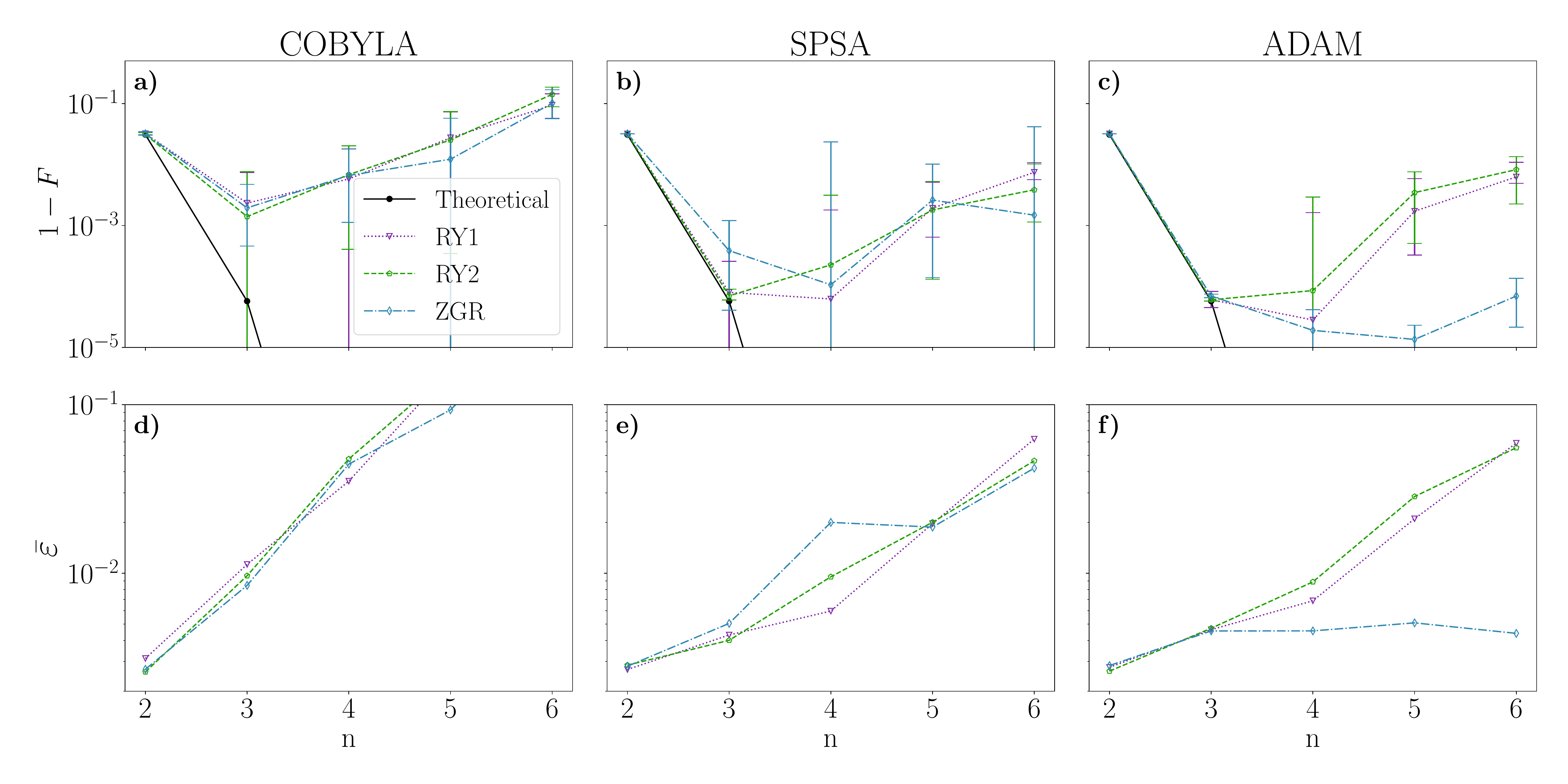}
\caption{Results of the simulations for two, three, four, five, and six qubits with 8192 evaluations for the harmonic oscillator using the ZGR and the $R_Y$ \textit{Ansätze} with depths 1 (RY1) and 2 (RY2) and the COBYLA, SPSA and ADAM optimizers. (a) Continuous infidelity with $n+m=12$. (b) Rescaled energy $\bar{\varepsilon}$ \eqref{eq: eps}.}
\label{Fig:results_ho}
\end{figure*}

\begin{figure}[hbt!]
\centering
\includegraphics[width=0.5\textwidth]{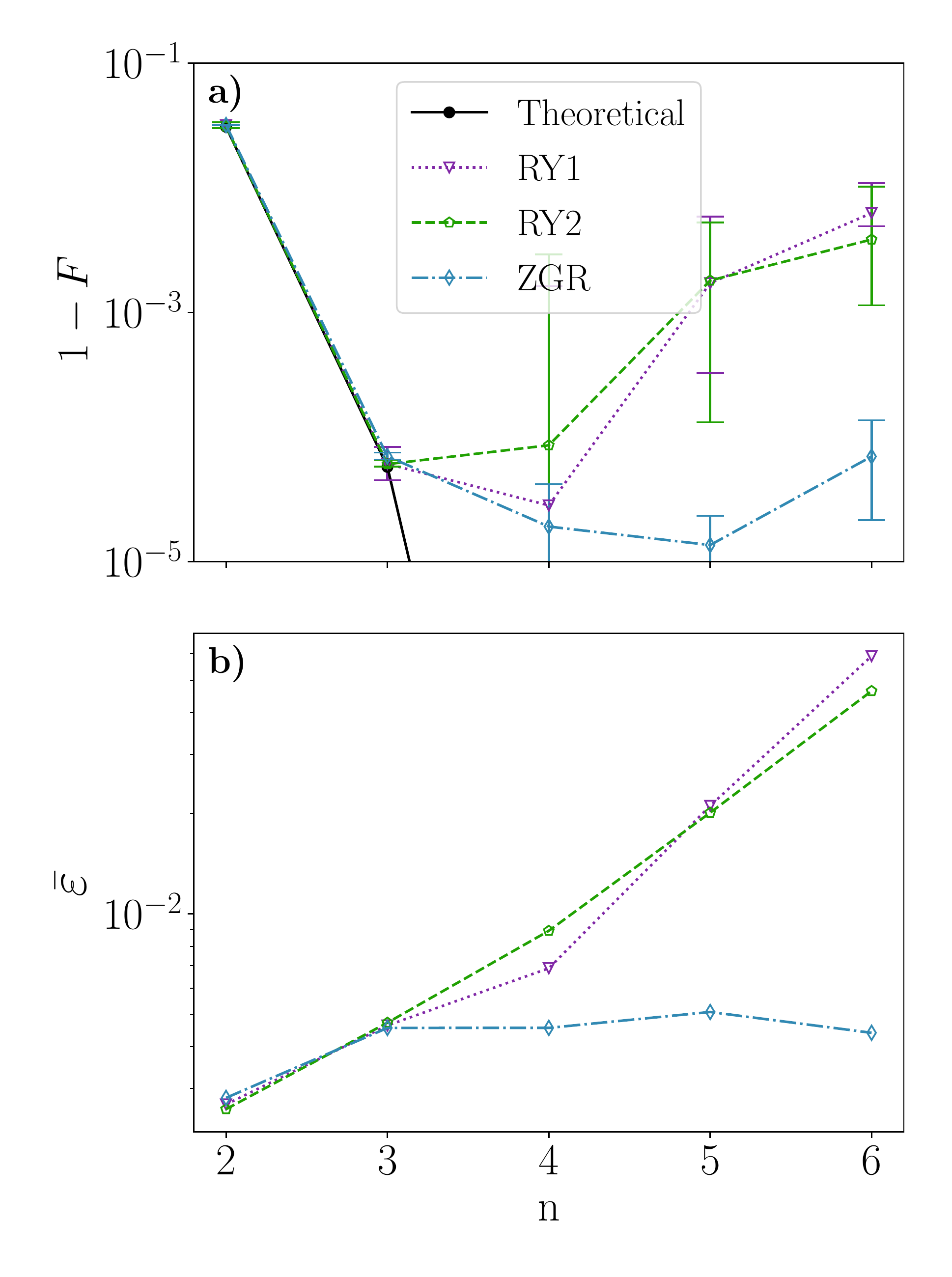}
\caption{Lowest infidelity results of the simulations for two, three, four, five, and six qubits with 8192 evaluations for the harmonic oscillator using the ZGR and the $R_Y$ \textit{Ansätze} with depths 1 (RY1) and 2 (RY2). (a) Continuous infidelity with $n+m=12$. (b) Rescaled energy $\bar{\varepsilon}$ \eqref{eq: eps}.}
\label{Fig:best_ho}
\end{figure}

\begin{figure}[h]
\centering
\includegraphics[width=0.5\textwidth]{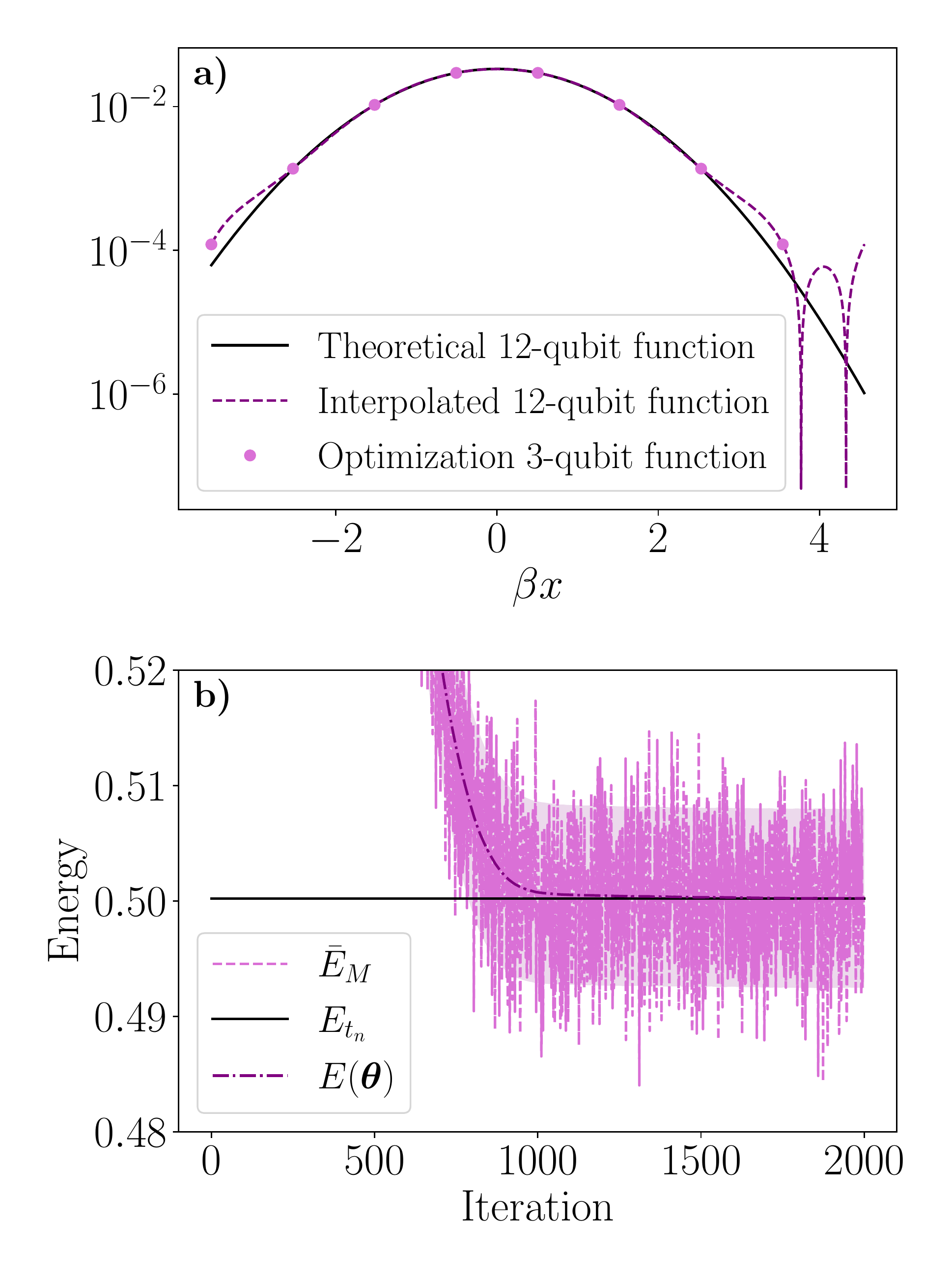}
\caption{Result of the optimization for the Adam optimizer and the ZGR \textit{Ansatz} for three qubits for the harmonic oscillator. a) Absolute value of the theoretical and optimization wavefunctions ($\beta x$ is a dimensionless coordinate where $\beta = \sqrt{m\omega/\hbar}$). b) Value of the energy for each iteration.}
\label{Fig:wavefunction}
\end{figure}

\begin{figure}[hbt!]
\centering
\includegraphics[width=0.5\textwidth]{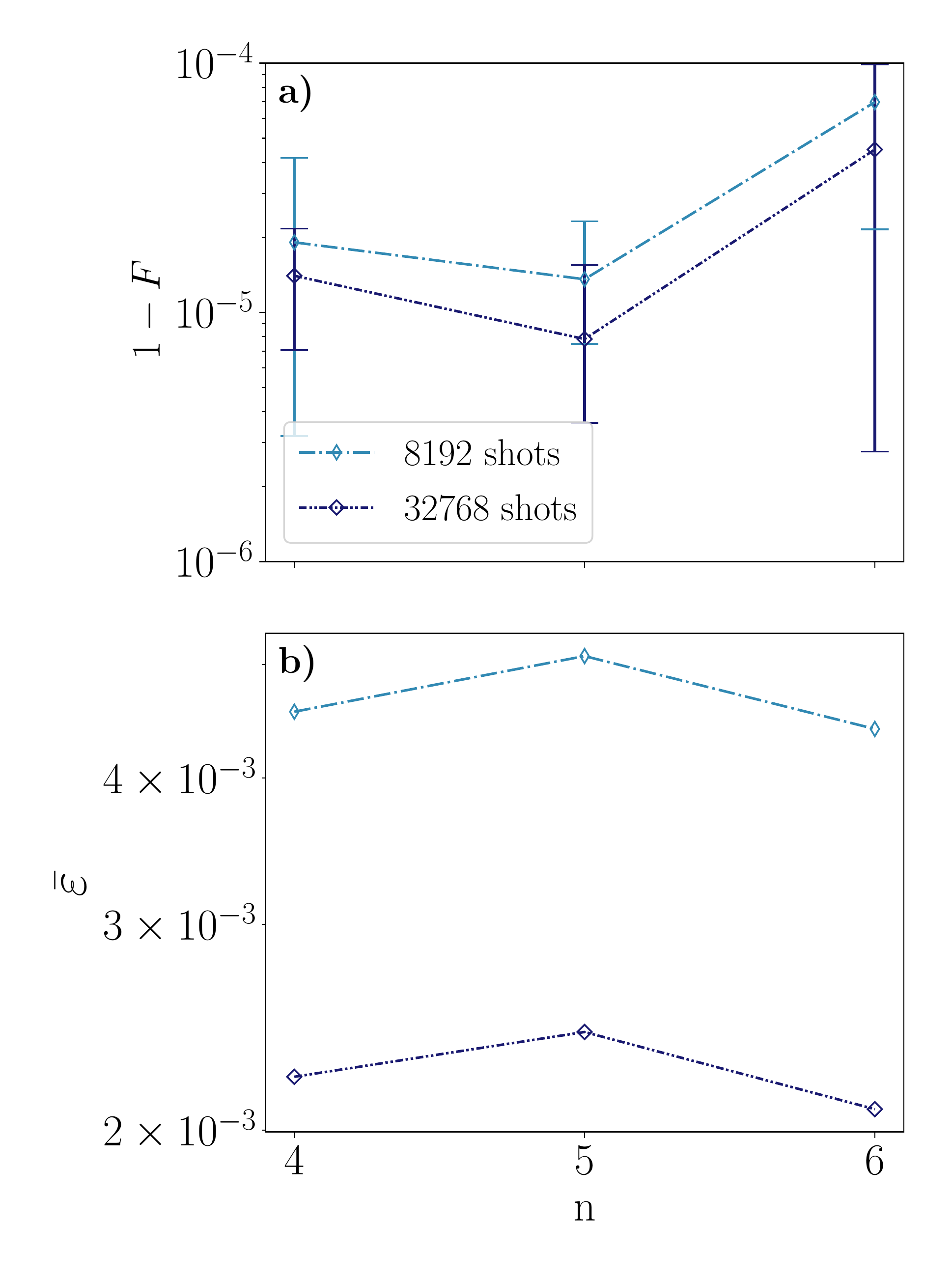}
\caption{Result of the optimization for the Adam optimizer and the ZGR \textit{Ansatz} for 8192 and 32768 evaluations for the harmonic oscillator. (a) Continuous infidelity with $n+m=12$. (b) Rescaled energy $\bar{\varepsilon}$ \eqref{eq: eps}.}
\label{Fig:32768}
\end{figure}

\subsection{Figures of merit}

When solving the PDEs we can compare the performance of the algorithms using different metrics, such as time, number of evaluations of the cost function, precision in the evaluation of the energy and precision in the determination of the encoded function. We focus on the last two.

To analyze the quality of the variational state at the end of the simulation we use the fidelity $F[\psi_1,\psi_2]:=|\langle{\psi_1|\psi_2}\rangle|^2.$ We may compute the fidelity between the variational states produced by the optimization $W(\bm\theta_\text{opt})\ket{0},$ with the discretized function using the same number of qubits $\ket{f^{(n)}}$
\begin{equation}
  F^{(n)}:=\left|\langle{f^{(n)}|W(\bm\theta_\text{opt})|0^{\otimes n}}\rangle\right|^2.
\end{equation}
This first figure of merit may be arbitrarily small, but it does not characterize how much information we have about the continuous function that is the solution to our problem $f(x).$ To fully understand this we need to gauge the quality of the continuous function that we associate with the quantum register~\eqref{eq:continuous-position}. This gives us a different figure of merit, which we call the continuous fidelity, which is obtained by using the quantum Fourier interpolation algorithm~\eqref{eq:interpolation}
\begin{equation} F^{\infty}:=\lim_{m\to\infty}\left|\langle{f^{(n+m)}|U_\text{int}^{n,m}W(\bm\theta_\text{opt})|0^{\otimes n}}\rangle\right|^2.
\end{equation}

In Fig.\ \ref{Fig:infid_comparison} we compare the infidelities $1-F^{(n)}$ and $1-F^{\infty}$ obtained with the ZGR \textit{Ansatz} and the best optimization method. We also show, for comparison, the best numerical approximation~\eqref{eq:continuous-position} that we can obtain using a finite grid with $2^n$ points. Note how the errors in the quantum state overestimate our true knowledge of the function $F^{(n)}\geqslant	 F^{\infty}.$ Consequently, in our later plots we will represent the median of the infidelities $1-F^{\infty}$ over 100 repetitions of each simulation with different initial states or trajectories, and the error bars will be the standard deviation around the mean. Moreover, we estimate $F^{\infty}$ using $m=12-n$ extra qubits, which already gives a good converged measure.

In addition to the quantum state, we are also interested in how well we can estimate the properties of the solution to the PDE. We gauge this by evaluating the relative error in the computation of the energy
\begin{equation} \label{eq: eps}
  \varepsilon = \left| \frac{E_{{t}_n}-E_\text{opt}}{E_1 - E_0} \right|.
\end{equation}
This is a dimensionless figure of merit that is adapted to the natural energy scales of the problem: $E_{{t}_n}$ is the theoretical energy obtained over a grid with $2^n$ points, $E_\text{opt}$ is the optimal energy derived by the algorithm, and $E_1-E_0$ is the energy difference between the lowest and first excited solutions of the PDE\ \eqref{eq:pde}. In idealized applications we expect $\varepsilon$ and $1-F^\infty$ to be proportional to each other, but this is not always true in real-world quantum computers, as we see below.

\subsection{Harmonic oscillator}
\label{sec:numerics-ho}

We applied the variational quantum PDE solver to the harmonic oscillator model from Sec.~\ref{subsec: ho}. The actual results are summarized in Table\ \ref{tab:ho} and Fig.\ \ref{Fig:results_ho}. Figures\ \ref{Fig:results_ho}(a)-(c) show the infidelity $1-F^\infty$ of the optimal state, as obtained with three different optimizers and three versions of the variational \textit{Ansatz}. As reference, we also plot the lowest infidelity obtained with the best approximation\ \eqref{eq:continuous-position} on the same grid. Figures\ \ref{Fig:results_ho}(d)-(f) illustrate the relative error in the prediction of the energy~\eqref{eq: eps} for the same optimal states.

From these figures we conclude that the best optimization method is Adam, closely followed by SPSA. Both methods seem to excel due to their tolerance to the intrinsic uncertainty in the estimation of the energy. However, while Adam uses an analytic estimate of the cost-function's gradient\ \cite{Mitarai2018, Schuld2019}, SPSA relies on an stochastic estimate with errors that get amplified by small denominators. Thus, even for the same \textit{Ansatz}, SPSA leads to worse estimates of the function and of the optimal energy.

From Fig.\ \ref{Fig:best_ho} we conclude that the ZGR \textit{Ansatz} is the best variational \textit{Ansatz} for this problem in the limited number of measurements. We attribute the difference in precision to the fact that the ZGR \textit{Ansatz} is designed for the representation of continuous functions\ \cite{Zalka1998, GroverRudolph2002}. In this \textit{Ansatz}, every rotation builds on the previous ones in a smooth, easily differentiable fashion, without any loss of information. In the $R_Y$ \textit{Ansatz}, however, the influence of different qubits and layers is more inefficiently transported by the layers of entangling unitaries, leading to the vanishing of gradients\ \cite{McClean2018}. This chaotic nature is more manifest as we increase the number of qubits and parameters to optimize.

Also in Fig.\ \ref{Fig:best_ho} we see that the number of measurements limits the achievable precision in any of the \textit{Ansätze} (compare with Table\ \ref{tab:ho}, L-BFGS-B column). However, the infidelities obtained, in the range $10^{-3}-10^{-5}$, are surprisingly below what is expected from the statistical uncertainty with which we evaluate the cost function. To illustrate this, Fig.\ \ref{Fig:wavefunction}(b) shows one of the stochastic trajectories created by the Adam method for the ZGR \textit{Ansatz}. As a dashed line we plot the evaluation of the cost function as returned by the simulator $\bar{E}_M(\bm\theta)$, surrounded by a colored band that estimates the statistical uncertainty for $M=8192$ measurements. This trajectory and error band must be compared to the actual energy computed for the same parameters $E(\bm\theta)$, without any uncertainty. This quantity approaches a relative error $\bar{\varepsilon}\simeq 5\times10^{-3},$ which is one order of magnitude below the statistical uncertainty, illustrating the power of stochastic optimization. Not only is the energy very well approximated by the stochastic optimization, but as shown in Fig.\ \ref{Fig:wavefunction}(a) we observe how with just three qubits we recover the theoretical wavefunction with high fidelity using interpolation. Very small Gibbs oscillations appear in the boundary of the interpolation interval due to not strictly periodic conditions.

With Fig.\ \ref{Fig:32768} we explore the behavior of the algorithm as the number of evaluations increases. We show that the precision of the algorithm is limited by the number of evaluations in the quantum computer, as the more evaluations the better the estimate of the expectation value [Fig. \ref{Fig:32768}(b)]. By increasing the number of evaluations by a factor 4, we decrease the value of $\bar{\varepsilon}$ by a factor 2. The increase in the precision of the algorithm is also manifest in the increase of the fidelity [Fig.\ \ref{Fig:32768}(a)].

\subsection{Transmon qubit}
\label{sec:numerics-transmon}

\begin{figure*}[b!]
\centering
\includegraphics[width=1\textwidth]{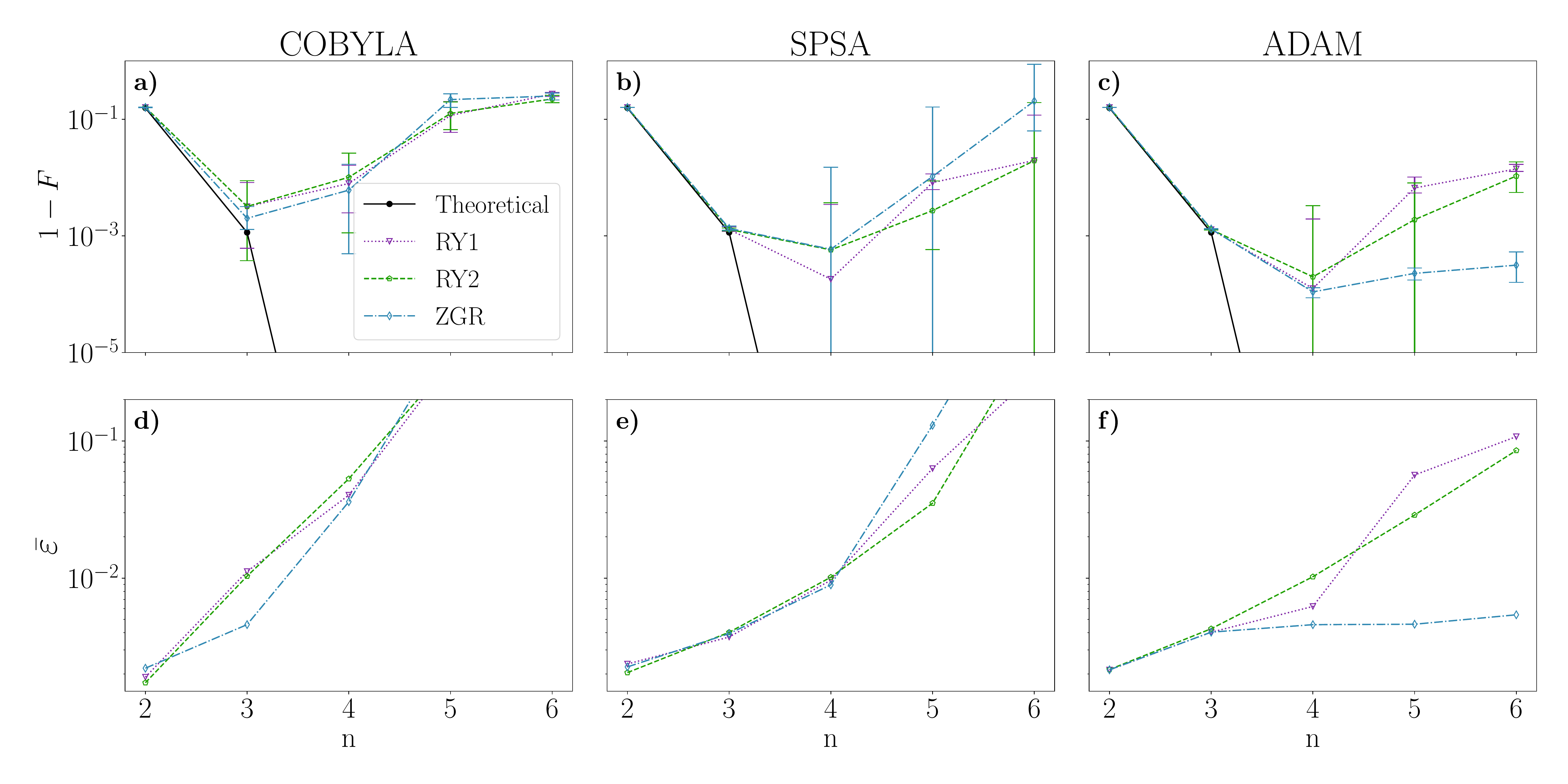}
\caption{Results of the simulations for two, three, four, five, and six qubits with 8192 evaluations for the transmon qubit using the ZGR and the $R_Y$ \textit{Ansätze} with depths 1 (RY1) and 2 (RY2) and the COBYLA, SPSA and ADAM optimizers. (a) Continuous infidelity with $n+m=12$. (b) Rescaled energy $\bar{\varepsilon}$ \eqref{eq: eps}.}
\label{Fig:results_transmon}
\end{figure*}

The results of the transmon qubit simulations are shown in Figs.\ \ref{Fig:results_transmon}(a)-(f). These plots confirm the observations made for the harmonic oscillator. Once more, the Adam optimizer leads to the lowest infidelity results, due to the use of the analytic gradient. The ZGR \textit{Ansatz} also behaves better than the $R_Y$ \textit{Ansatz}, and no \textit{Ansatz} achieves the minimum infidelity that can be reached with three to six qubits, probably for the same reasons as before: local trapping and statistical uncertainty.

\begin{figure*}
\centering
\includegraphics[width=1\textwidth]{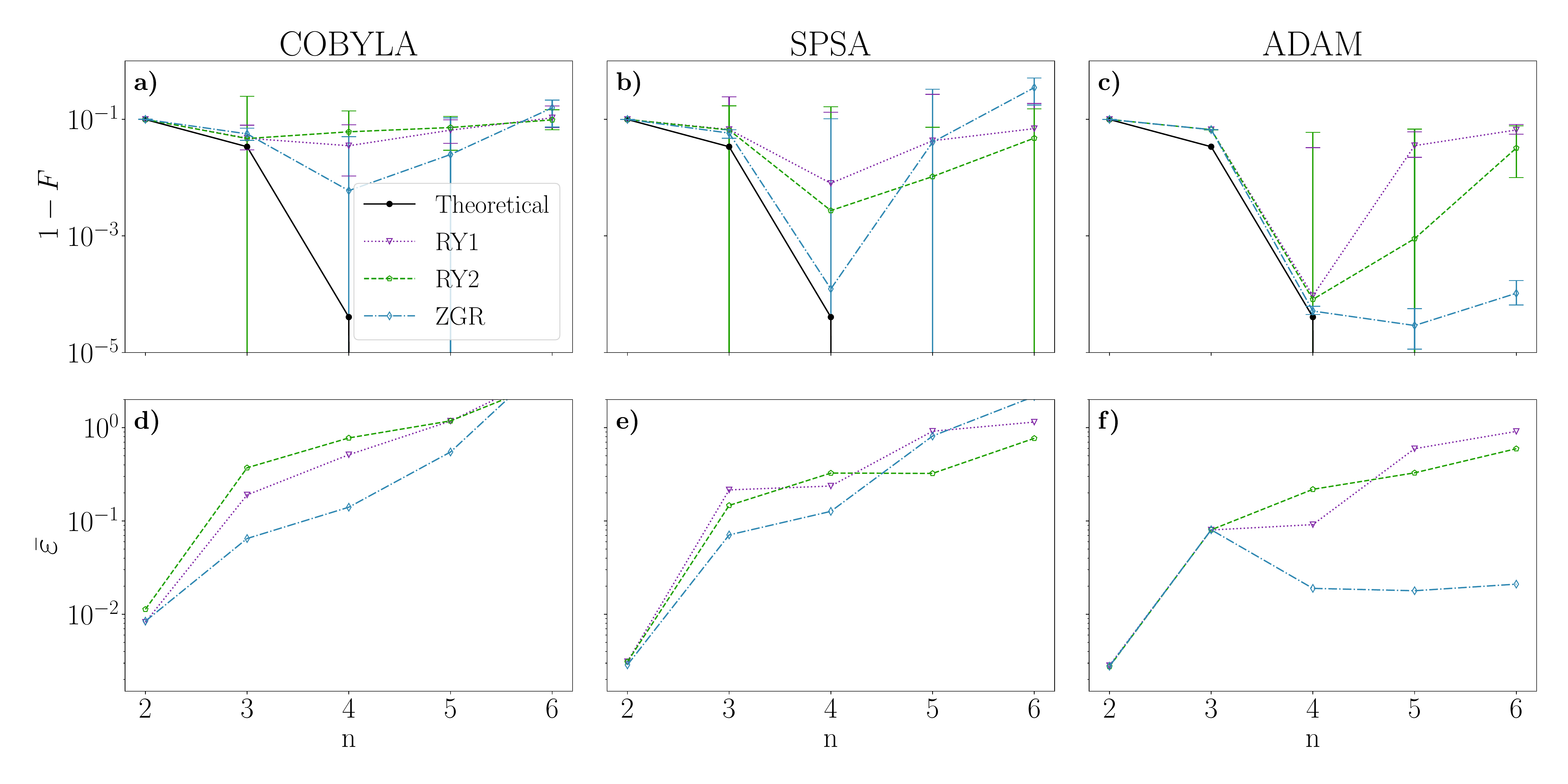}
\caption{Results of the simulations for two, three, four, five, and six qubits with 8192 evaluations for the flux qubit using the ZGR and the $R_Y$ \textit{Ansätze} with depths 1 (RY1) and 2 (RY2) and the COBYLA, SPSA and ADAM optimizers. (a) Continuous infidelity with $n+m=12$. (b) Rescaled energy $\bar{\varepsilon}$ \eqref{eq: eps}.}
\label{Fig:results_flux_qubit}
\end{figure*}

When we compare the harmonic oscillator and the transmon, we see that the latter is typically affected by greater infidelities. The zeroth-order solution of the Mathieu equation\ \eqref{eq:eq_transmon}, which is the transmon ground state, is more complicated to reproduce than the Gaussian function of the ground state of the harmonic oscillator\ \eqref{eq:eq_ho}. First, this is a periodic function that does not strictly vanish on the boundaries. Second, contrary to the Gaussian function, we are not allowed to change the length of the interval to maximize the precision in both position and momentum space. All together, as shown in Table\ \ref{tab:transmon}, the achievable infidelities are in general worse for all \textit{Ansätze}, even when using numerical exact optimizations (L-BFGS-B).

We chose the transmon qubit equation because it is a physically motivated problem. One might wonder about the utility of this method for the computation of actual properties, such as the energies and excitation probabilities of actual qubits. The relative errors that we have obtained, $10^{-2}-10^{-3}$, are compatible with what can be expected from using 8192 shots. One could further decrease to the theoretical limits, using more measurements. However, in order to achieve a relative error below $10^{-4}$ (a fraction of a megahertz), one would need to use about 100-10000 times more measurements in the final stages of the optimization. While this seems doable, it suggests the need to find better strategies for the energy evaluation or even the optimization itself.

\subsection{Flux qubit}
\label{sec:numerics-flux_qubit}

We show the results of the resolution of the flux qubit equation \eqref{eq: flux qubit} in Figs.\ \ref{Fig:results_flux_qubit}(a)-(f) and Table \ref{tab:flux qubit}. The best performance is obtained for the ZGR \textit{Ansatz} combined with the Adam optimizer. These results corroborate our previous observations, even though the ground state of the flux qubit is not a Gaussian and cannot by approximated by one. Thus, we can affirm that our method has succeeded at obtaining different types of solutions, as long as they verify the conditions in Sec.\ \ref{sec:solver}. 
\begin{figure}
\centering
\includegraphics[width=0.5\textwidth]{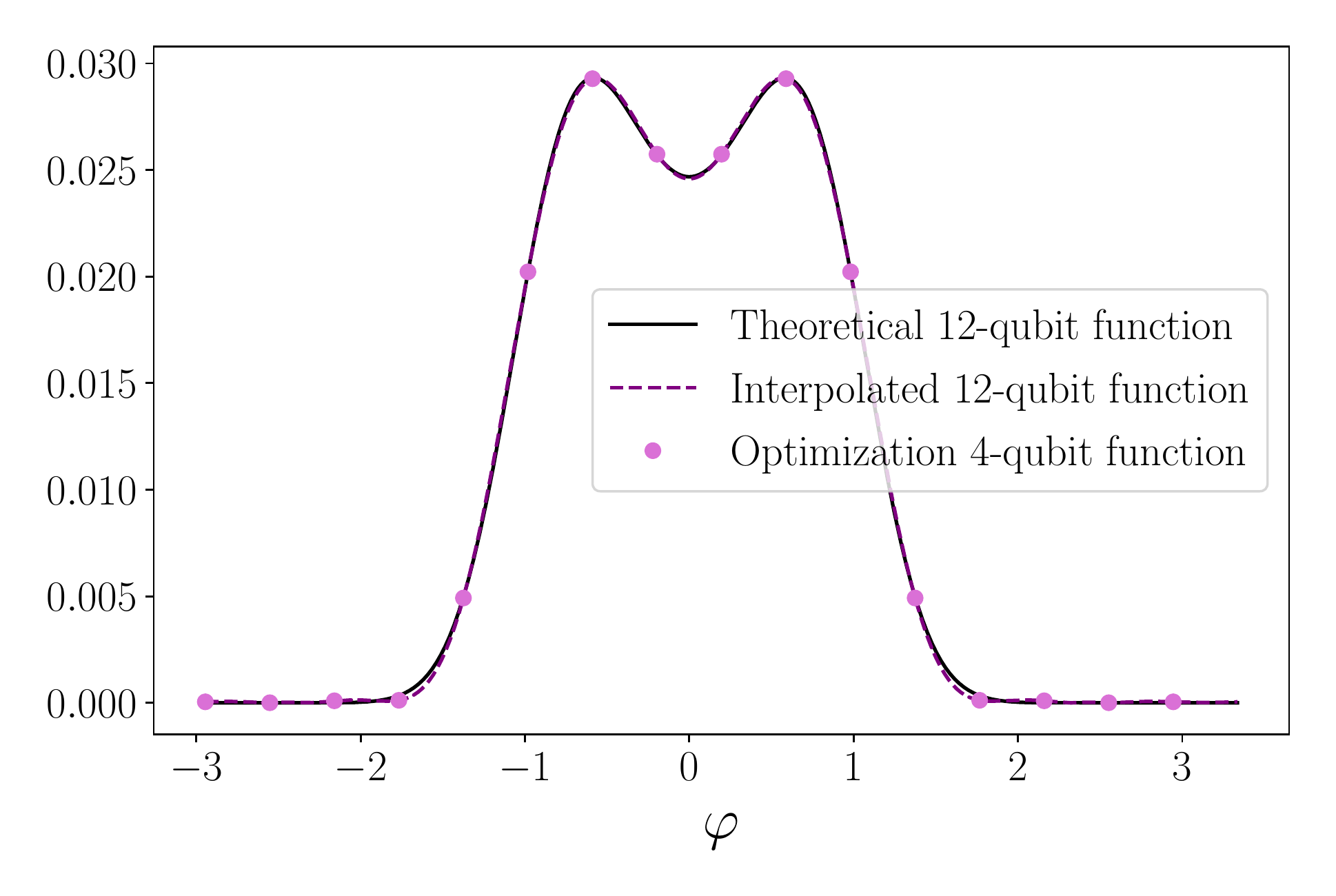}
\caption{Absolute value of the ground state of the flux qubit. We show the theoretical eigenstate as well as the discretized and interpolated solution of our algorithm for the four-qubit RY1 \textit{Ansatz} using the L-BFGS-B optimizer.}
\label{Fig: flux qubit ground state}
\end{figure}
The flux qubit ground state has no analytic solution, but we can compare the ground state to the numerically exact solution of the PDE solved by alternative techniques, in this case, the representation of the qubit in the charge basis. As shown in Fig.\ \ref{Fig: flux qubit ground state}, the double-well structure of the qubit's potential creates a state that looks like the superposition of two Gaussians.
The greater complexity of the flux qubit's ground state demands more qubits for an accurate representation, but already four qubits is enough to reach good results ($10^{-5}$ infidelity) for the function representation. In contrast, the relative errors for the energy of the qubit are higher, but this can be explained by the fact that the ground and first excited state energies are comparatively closer, requiring a smaller absolute precision in the evaluation of the energy to be approximated.
\color{black}
\subsection{Application to NISQ devices}
\label{sec:numerics-errors}

Until now we have focused on the performance of our algorithm under noiseless circumstances. However, the actual motivation of variational methods is to work in NISQ devices, where qubits have finite lifetimes and gates are imperfect, but even imperfect variational constructs can be optimized to approach the ideal limits\ \cite{Sharma2020}. In a similar spirit, we will now evaluate the noise tolerance of the variational quantum PDE solver, to understand its suitability for existing and near-term quantum computers.
\begin{figure}
\centering
\includegraphics[width=0.5\textwidth]{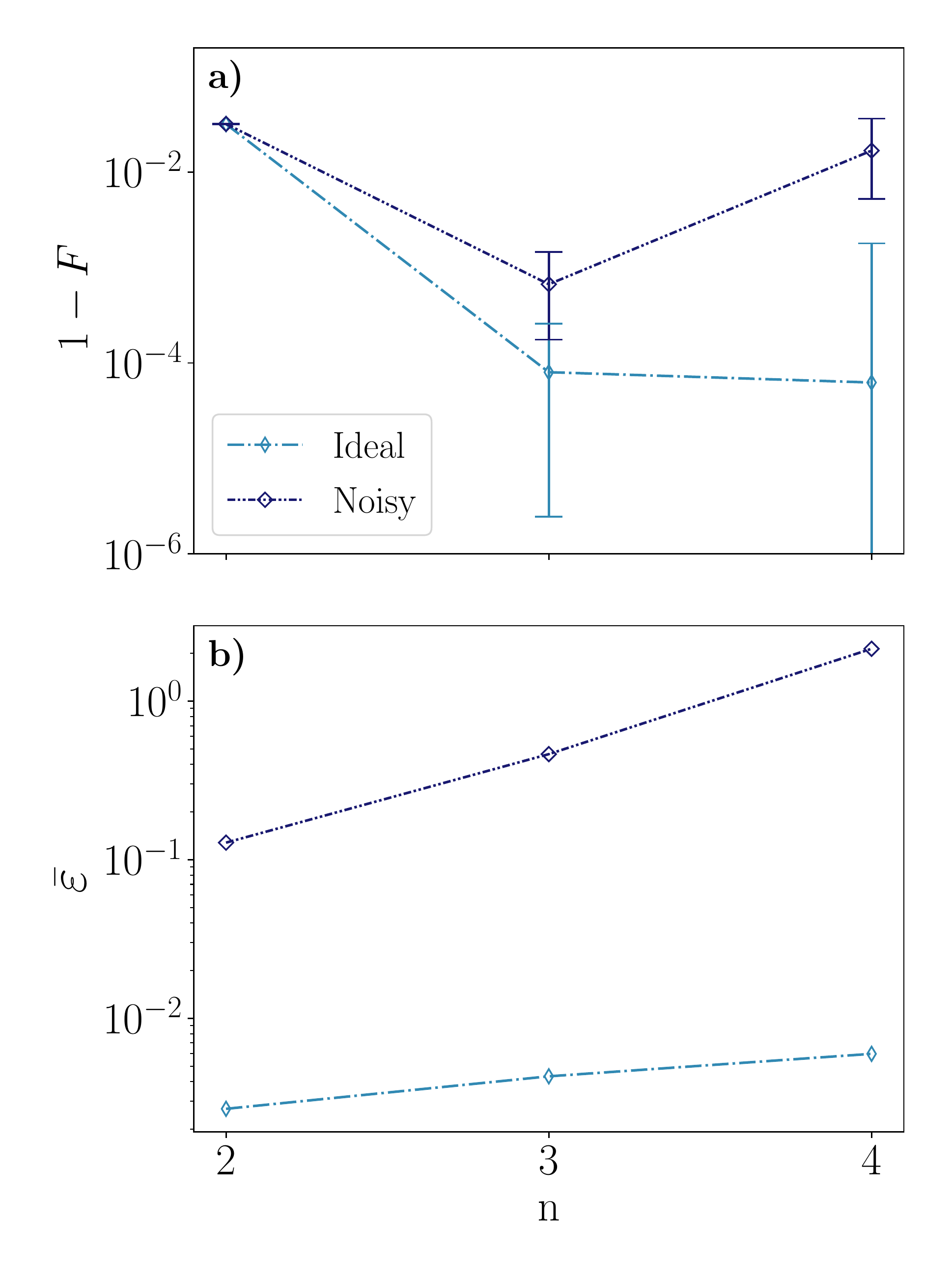}
\caption{Results for the ideal (noiseless) and noisy (ibmq\_santiago noise model) optimization for the SPSA optimizer and the RY1 \textit{Ansatz} with 8192 evaluations for the harmonic oscillator. (a) Continuous infidelity with $n+m=12$. (b) Rescaled energy $\bar{\varepsilon}$ \eqref{eq: eps}.}
\label{Fig:idealvsnoisy}
\end{figure}
Our study is performed using a simulator of the ibmq\_santiago five-qubit   quantum computer\ \cite{Qiskit}. This simulator allows us to store the noise model, coupling map, and basis gates so that each computation is subject to identical conditions. The noise model includes the gate error probability and gate length for each basis gate and qubit, and the readout error probabilities and $T_1$ and $T_2$ relaxation times for each qubit. For this device, $T_1$ and $T_2$ are of the order of $100\ \mu$s, and the readout probability is $\sim 10^{-2}$. The gate error for single-qubit gates is of order $10^{-3}-10^{-4}$ with gate length of order $10$ ns, while both the gate error and length are one order of magnitude greater for two-qubit gates- \footnote{The results are for the ibmq\_santiago calibration with date 17 March 2021.}

Our study applies the $R_Y$ \textit{Ansatz} with depth 1 to the solution of the harmonic oscillator, using the SPSA optimizer. We chose the \textit{Ansatz} and problem that provide the best fidelities, with the least number of gates, combined with an optimizer that is expected to perform well in noisy problems. Figure\ \ref{Fig:idealvsnoisy}(a) shows the infidelity of the continuous function in the noiseless and noisy simulations. The fidelity decreases with the number of qubits due to the greater number of gates, which introduce more errors and increase the effect of decoherence due to the longer time of the circuit. However, for a small number of qubits, it is possible to obtain a reasonably small infidelity of order $10^{-4}$ for three qubits, which allows us to reconstruct the theoretical continuous wave function.

\begin{figure}
\centering
\includegraphics[width=0.5\textwidth]{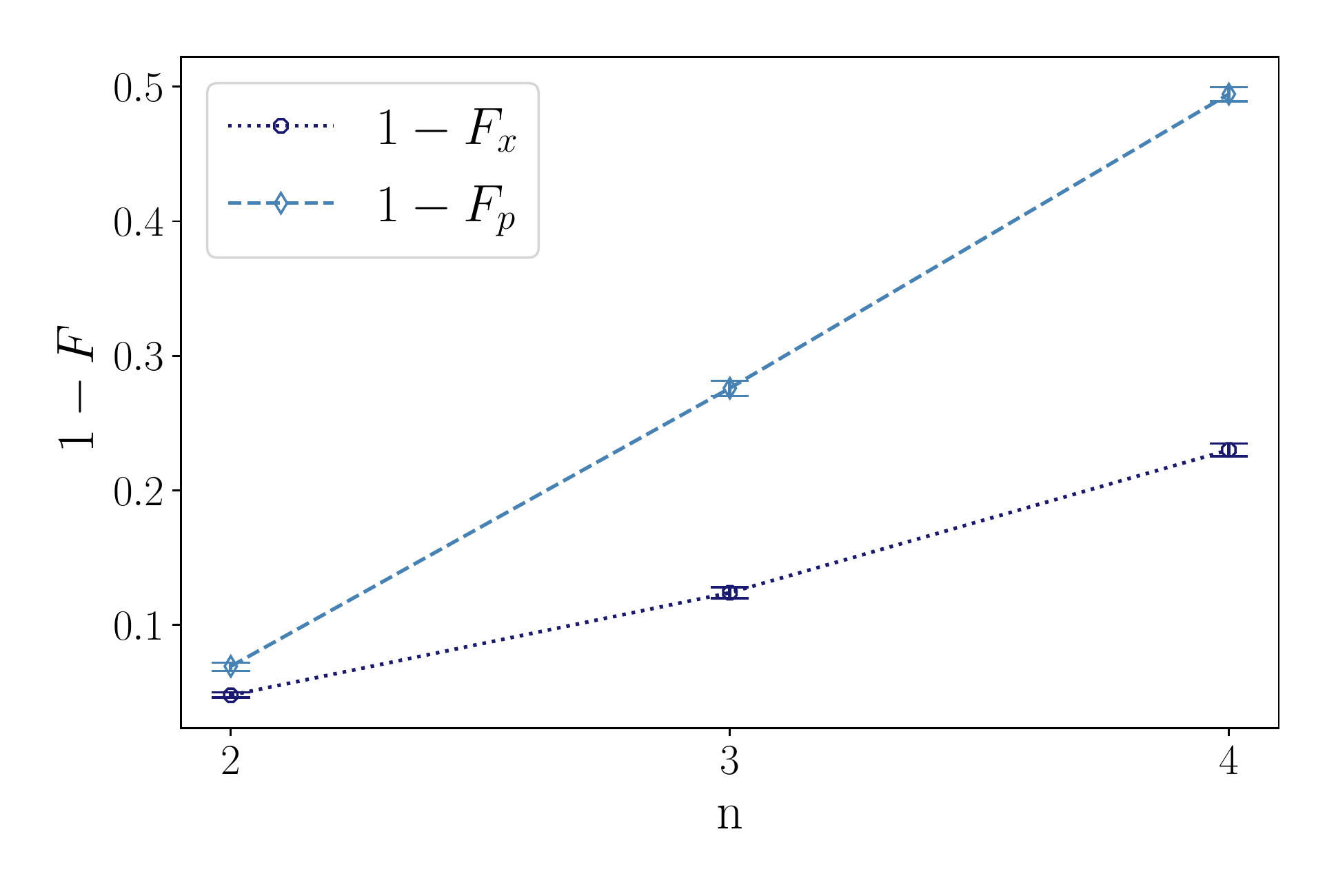}
\caption{Circuit infidelity for the position $1-F_x$ and momentum $1-F_p$ circuits for the harmonic oscillator for the RY1 \textit{Ansatz} for $100$ repetitions with $8192$ evaluations each and using the ibmq\_santiago noise model.}
\label{Fig:tomo}
\end{figure}

Although the optimization of the parameters is successful, we obtain a very significant error in the evaluation of the energy [Fig.\ \ref{Fig:idealvsnoisy}(b)], even for a small number of qubits. We attribute this error to the circuits that we use to evaluate the energy in position and momentum space. We can confirm this hypothesis using the QISKIT quantum tomography toolbox for  a small number of qubits, using the ibmq\_santiago simulator. As shown in Fig.\ \ref{Fig:tomo},  the errors in the two circuits can be quite significant and are larger in momentum space because of the gates that are required for the QFT.

\begin{figure}
\centering
\includegraphics[width=0.5\textwidth]{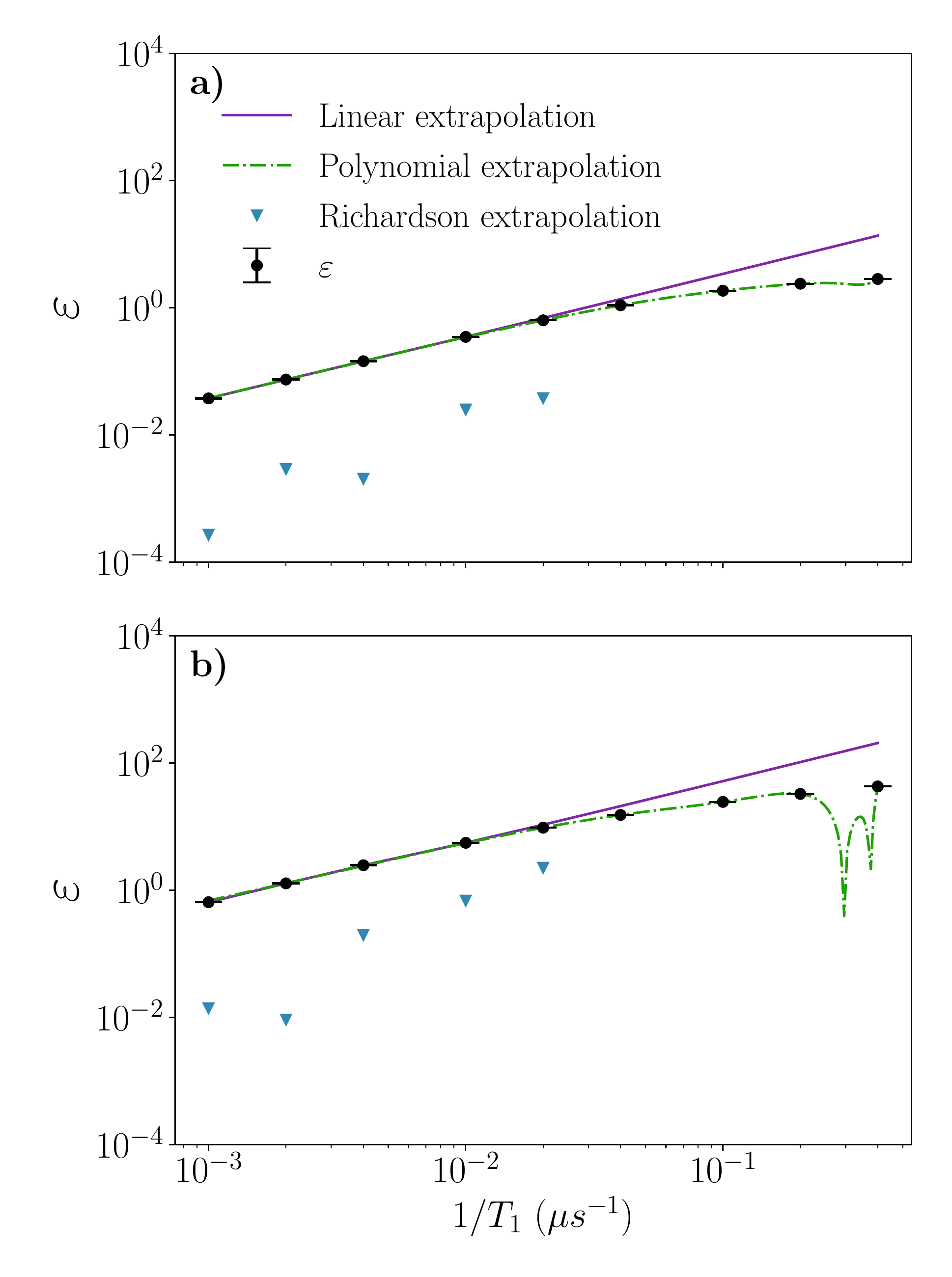}
\caption{Zero-noise extrapolation results (RY1 \textit{Ansatz} with thermal relaxation and $100$ repetitions with $8192$ evaluations for each simulation). (a) Harmonic oscillator (three qubits). (b) Flux qubit (four qubits).}
\label{Fig:noisyenergy}
\end{figure}

Despite the errors in these circuits, the fact that we obtain a good quality estimate of the function itself suggests that we can apply error mitigation and zero-noise extrapolation\ \cite{Kandala2019}. To test this hypothesis, we repeat the simulations using a simpler noise model dominated by thermal relaxation, where we can tune $T_1,$ but still use the coupling map and basis gates of the ibmq\_santiago five-qubit device. In this study we prepare a quantum circuit with the optimal solution for three qubits and compute the mean energy averaging over $100$ repetitions of the simulation with $8192$ evaluations each of them, for different values of $T_1.$ We use this mean energy to compute $\varepsilon$ [Eq.\ \eqref{eq: eps}]. As Fig.\ \ref{Fig:noisyenergy}(a) shows, if $T_1\geqslant	 2.5\mu\mbox{s},$ the energy admits a Taylor expansion of fifth order
\begin{equation}
  E(T_1) = E_0 + \sum_n \epsilon_n \frac{1}{T_1^n}.
\end{equation}
This expansion can be used by itself or with Richardson extrapolation\ \cite{Kandala2019} to estimate values of the energy with a lower error,  of order $\varepsilon \sim 10^{-2}$ for values of $T_1 \sim 50-100 \ \mu s$, which are within the experimental range.

{We have also tested the noise resilience of our algorithm using a more complex equation, the Schrödinger equation of the flux qubit~\eqref{eq: flux qubit}. Let us recall that the energy spectrum of the flux qubit has a much smaller gap between the ground and first excited states than the one of the transmon. Moreover, the wave function is a superposition of two Gaussians on two separate wells and thus requires more points to achieve a low-infidelity interpolation. Overall, this results in greater errors in the estimation of the energy, even under noiseless circumstances 
(c.f. Figs.~\ref{Fig:results_flux_qubit}). Therefore, the error mitigation is expected to perform worse than for the harmonic oscillator.

In Fig.~\ref{Fig:noisyenergy}(b) we show the results of the noisy simulations with four qubits for the flux qubit. We observe that the increase of the number of qubits, and hence quantum gates and their associated errors, together with the nature of this equation, leads to worse results, as expected. The values of $\varepsilon$ admit a fifth-order Taylor expansion, only if $T_1\geqslant	 5\mu$s. Moreover, when using Richardson extrapolation, we cannot obtain a low error in the estimation of the energy for state-of-the-art values of $T_1$, and just for values of $500-1000 \mu$s we reach errors of order $\varepsilon \sim 10^{-2}$. This points to the necessity of larger thermal relaxation times in real noisy quantum computers to estimate the ground-state energy of more complex Hamiltonians.}

\section{Conclusions and future perspectives}
\label{sec:conclusion}

{We have developed a set of quantum Fourier analysis techniques to efficiently encode continuous functions and differential operators in quantum registers. Our study showed that spectral methods leverage the potential of the quantum register, approximating these problems with an error that decreases doubly exponentially in the number of qubits, in optimal circumstances. We illustrated the performance of these methods by creating a variational quantum algorithm to solve the eigenvalue problem of physically motivated} Hamiltonian PDEs. We combined this algorithm with newer variational \textit{Ansätze} that are better suited to describe continuous differential functions or which include symmetries. We have tested these ideas in idealized scenarios with infinite or a limited number of measurements, obtaining excellent accuracy ($10^{-5}$ infidelity) even with small numbers of qubits. In addition, under the presence of noise sources, the algorithm is still efficient, reaching high-fidelity results ($10^{-3}-10^{-4}$ infidelity) for three qubits. 

As in all variational methods, we have seen that the accuracy is ultimately limited by our statistical uncertainty in the determination of the cost function and the analytic gradients. Assume, for instance, that we want to use the present method to study a transmon qubit. Reaching the required experimental precision, with errors $\varepsilon\simeq 10^{-4}$, can be costly in a scenario of NISQ computers with limited access and temporal stability. It also requires significant improvements in both the classical optimization and the quantum evaluation of properties. These are the limitations that current NISQ devices need to overcome to achieve the favorable scalings provided by our quantum Fourier algorithms. In such a scenario, it may be worth considering alternative quantum-inspired methods that can benefit from similar encodings, and also provide heuristic advantages for the same type of equations\ \cite{GarciaRipoll2021}. Nevertheless, our algorithm succeeds in proving the high efficiency of the quantum Fourier analysis representation, achieving high fidelities with a low number of qubits (three to four qubits).

{More generally, our investigation opens several avenues for future research. First, less demanding gradient computing techniques can also be implemented\ \cite{Sweke2020}, which will also be required to extend this method to solve  in higher dimensional spaces. Second, the good results obtained for the Ry \textit{Ansatz} with low depth, as well as with the more general ZGR \textit{Ansatz}, suggest the possibility to develop simplified \textit{Ansätze} that encode smooth functions in quantum states with lower circuit depth. Such a possibility is supported by heuristic studies of the entropy growth with the discretization density\ \cite{GarciaRipoll2021}, but a precise scaling and the ideal structure of the circuit is still missing. Undoubtedly, such results would also be of interest for applications in quantum machine learning.

Regarding the generality of the equations, our work has focused on Hermitian operators with lower bounded spectra. For more general static equations, which are still of the form $(D(\nabla) + V - E)f=0,$ one may extend the algorithm to work with self-adjoint operators, as it is usually done in the study of Lindblad operators, by analyzing the square of the equation $(D(\nabla) + V - E)^\dagger(D(\nabla) + V - E)f=0.$ The main obstacle in this analysis is the cost of studying products such as $D(\nabla) V(\mathbf{x}),$ which, while still amenable to computation with our methods, undoubtedly demand a larger number of gates and probably must be delegated to error-corrected quantum computers.

With respect to fault-tolerant quantum computers, we see many opportunities to enlarge the scope of quantum numerical analysis algorithms based on spectral methods. As discussed in the text, Fourier encodings have favorable scaling in the number of qubits needed to encode smooth, highly differentiable functions, such as the ones we find in many relevant physical problems, and also in other realms of scientific computing. One interesting avenue for study is to combine spectral methods with the tools of quantum simulation to solve initial-value problems, which are more challenging than the eigenvalue problems developed here. Moreover, it will be interesting to explore how these methods can be merged with amplitude estimation to better interrogate the properties of general functions.}

\section{Acknowledgments}

This work was funded by MCIN/AEI/10.13039/501100011033 and "FSE invierte en tu futuro" through an FPU Grant FPU19/03590. J.J.G.-R. and P.G.-M. acknowledge support from Spanish Project No. PGC2018-094792-B-I00 (MCIU/AEI/FEDER, UE), CAM/FEDER Project No.  S2018/TCS-4342 (QUITEMAD-CM), and CSIC Quantum Technology Platform PT-001.

\appendix
\twocolumngrid

\section*{APPENDIX}

\section*{A. FOURIER SPECTRAL METHOD ERRORS}

In this appendix we summarize well-known results from Fourier analysis regarding decomposition, truncation, interpolation, and differentiation of functions, following Refs.\ \cite{Kopriva2009,Canuto2006}. We use these results in the article to discuss the errors of our method.

\subsection{Fourier series errors}

Spectral methods approximate the solution of a differential equation using a sum of orthogonal basis functions \cite{Kopriva2009}. In this work we propose to use the Fourier modes. For simplicity, we can focus on one-dimensional function spaces, rescaled to the $[0,2\pi)$ interval. The Fourier modes
\begin{equation}
    \phi_k(x)=\frac{1}{\sqrt{2\pi}}e^{-ikx}, \quad k\in\mathbb{Z},
\end{equation}
are an orthonormal basis of square-integrable functions $L_{[0,2\pi]}^2.$ The Fourier series is the expansion of a function on this basis
\begin{equation}
    \label{eq: Fourier series}
    f(x) = \sum_{k=-\infty}^\infty \hat{f}_k \phi_k,
\end{equation}
using the Fourier coefficients $\hat{f}_k, \ k\in\mathbb{Z}$
\begin{equation}
    \label{eq: Fourier coefficients}
    \hat{f}_k = (\phi_k,f) = \frac{1}{\sqrt{2\pi}} \int_0^{2\pi} f(x) e^{ikx}dx.
\end{equation}

In numerical applications we can only take care of a limited number of Fourier coefficients, working with the truncated Fourier series
\begin{equation}\label{eq: truncated Fourier series}
    P_N f(x) = \sum_{k=-N/2}^{N/2} \hat{f}_k \phi_k.
\end{equation}
This truncation introduces an error that can be quantified as
\begin{equation}
  \Vert{f(x)-P_N f(x)}\Vert^2 = \sum_{|k|=N/2+1}^\infty |\hat{f}_k|^2.
\end{equation}
This error is finite, but it depends on how fast the Fourier coefficients \eqref{eq: Fourier coefficients} decay to zero, which they always do according to the Riemann-Lebesgue lemma. On the one hand, if $f(x)$ is $m$ times continuously differentiable in $[0,2\pi]$ ($m\geqslant	 1$) and if $f^{(j)}$ is periodic for all $j\leqslant m-2$, then $\hat{f}_k= O(1/k^m)$~\cite{Canuto2006}, and the truncation error for $f\in H_p^m(0,2\pi)$ (where $H_p^m$ is the Sobolev space, i.e., the space of functions whose first $m-1$ derivatives are periodic) is of order~\cite{Canuto2006}
\begin{equation}
    \Vert{f(x)-P_N f(x)}\Vert = O(N^{-m}).
\end{equation}
On the other hand, if $f(x)$ is an analytic function $f(x) \in C^\infty$ and periodic with all its derivatives on $[0,2\pi]$, the decay of the $k$th Fourier coefficient is faster than any negative power of $k$~\cite{Canuto2006}. More precisely, when $f(x)$ is periodic with period $2\pi$ and analytic in a strip of radius $r > 0$ centered around the real axis $|\text{Im}z < r|$, the truncation error decays exponentially~\cite{Canuto2006}
\begin{equation}
    \Vert{f(x)-P_N f(x)}\Vert = O(e^{-rN}).
\end{equation}

\subsection{Discrete interpolation errors}

Our method is based on an interpolation technique that replaces the Fourier coefficients in $P_Nf(x)$ with those coming from a discrete Fourier transform. Assuming a discretization of the interval $[0,2\pi)$ using $N$ points with  $x_j = 2\pi j / N$ and $-N/2 \leqslant k \leqslant N/2 -1,$ those coefficients are
\begin{equation}
    \label{eq: Discrete Fourier coefficients}
    \tilde{f}_k = \sqrt{\frac{2\pi}{N}}\sum_{j=0}^{N-1}f(x_j) \phi_j^* = \frac{1}{\sqrt{N}}\sum_{j=0}^{N-1}f(x_j)e^{ikx_j}.
\end{equation}
Using these coefficients we construct a Fourier interpolant $I_N f(x)$ of degree $N/2$ as the sum
\begin{equation}
    \label{eq: Discrete Fourier interpolation}
    I_N f(x) := \sqrt{\frac{2\pi}{N}}\sum_{k=-N/2}^{N/2-1}\tilde{f}_k \phi_k(x)
\end{equation}
that reproduces the values of the original function over the same lattice $I_Nf(x_j)=f(x_j).$

The discrete Fourier transform~\eqref{eq: Discrete Fourier coefficients} differs from the exact Fourier coefficients $\hat{f}_k$~[Eq.\ \eqref{eq: Fourier coefficients}] by an amount $R_N$ called the aliasing term
\begin{equation}
R_N := \tilde{f}_k - \hat{f}_k = \sum_{m=-\infty, m \neq 0}^\infty \hat{f}_{k\pm mN}.
\end{equation}
Due to the orthogonality of the Fourier coefficients, the Fourier series truncation and the aliasing onto the lattice commute. The difference between the original function and the interpolated one can thus be written as
\begin{align}
    \parallel
f(x)-I_N f(x)\parallel
^2 &= \parallel
f(x)-P_N f(x)\parallel
^2 \\
    &+\parallel
R_N f(x)\parallel
^2. \notag
\end{align}
The interpolation error $\parallel
f(x)-I_N f(x)\parallel
$ behaves asymptotically like the truncation error \cite{Kreiss1979, Canuto2006}, depending similarly on the differentiability properties of the function. In particular, for bandwidth-limited or more specifically analytical functions, the error of the interpolation decreases exponentially with the number of points on the lattice $N$.

\subsection{Gibbs phenomenon}

Even if the function under study is almost everywhere differentiable, a Fourier series will exhibit Gibbs oscillations around any discontinuity of the function or its derivatives. Those oscillations are evidence of the fact that such discontinuities require arbitrarily large frequencies. Assume for instance a piecewise continuously differentiable periodic function $f(x)$, $x\in[0,2\pi]$ with a jump discontinuity at $x=x_0$. Let us write its truncated Fourier series \cite{Canuto2006}
\begin{align}\label{eq: Dirichlet truncation}
    P_N f(x) &=\frac{1}{2\pi}\int_0^{2\pi}\left[\sum_{k=-N/2}^{N/2}e^{-ik(x-y)}\right]f(y)dy \\
    &= \frac{1}{2\pi}\int_0^{2\pi}D_N(x-y)f(y)dy, \notag
\end{align}
using the Dirichlet kernel $D_N(\xi)$
\begin{equation}
    D_N(\xi)=1+2\sum_{k=1}^{N/2}\cos\left(k\xi\right).
\end{equation}
Around the discontinuity, the Fourier series can be approximated~\cite{Canuto2006} by
\begin{align}
    P_N f(x) &\simeq \frac{1}{2}\left[f(x_0^+)+f(x_0^-)\right] \\
    &+ \frac{1}{2\pi}\left[f(x_0^+)-f(x_0^-)\right]\int_0^{x-x_0}D_N(y)dy, \notag
\end{align}
as $N\rightarrow\infty$. An analysis of the second term reveals that the Dirichlet kernel distorts the function's jump by a fixed factor $(0.089489872236\dots)$ that is around 9\%. Unfortunately, this phenomenon also affects the interpolated version of the function, as it can be seen by using the trapezoidal quadrature rule to relate~\eqref{eq: Dirichlet truncation} with
\begin{align} \label{eq: Dirichlet interpolation}
	I_N f(x) &= \frac{1}{N}\sum_{l=0}^{N-1}\sum_{k = -N/2}^{N/2} f(x_l) e^{-ik(x_j-x_l)} \\
	&= \frac{1}{N}\sum_{l=0}^{N-1} D_N(x_j-x_l) f(x_l). \notag
\end{align}
In practice, Gibbs oscillations only affect the precise values of the function around discontinuities. However, they do not affect the limit of computations with $P_Nf(x)$ or $I_Nf(x)$ in the approximations of integrals and other observables, where the effect of such distortions averages out and decays algebraically as $N\to\infty$.

\subsection{Differentiation errors}

We can use the exact Fourier transform to easily compute the derivative of a function $f(x)$. However, in our method we use a discretization of the derivative that results from differentiating the interpolated function $(D_Nf)(x):=\partial_x I_N f(x).$ As opposed to truncation, interpolation and differentiation do not commute due to the aliasing~\cite{Kopriva2009}.

We can analyze the errors associated with the interpolation derivative~\cite{Canuto2006} assuming a differentiable function $f\in H_p^m(0,2\pi)$ from a Sobolev space $H_p^m(0,2\pi)$ that supports $m\geqslant	 1$ derivatives
\begin{align} \label{eq: Differentiation error}
    \Vert{f'(x)-\mathcal{D}_N f(x)}\Vert = O(N^{1-m}). \notag
\end{align}
As before, if the function is analytic, we can improve this bound, which becomes exponentially small in the discretization~\cite{Tadmor1986}.

\newpage
\onecolumngrid
\subsection*{B. NUMERICAL RESULTS}

\begin{table}[H]
\centering
{\setlength\arrayrulewidth{0.2pt}
{\fontsize{2}{3}\selectfont\resizebox{\textwidth}{!}{\begin{tabular}{c c c c c c c c}\hline \hline
Qubits & \textit{Ansätze} & Parameters &  CNOT gates & \begin{tabular}[c]{@{}l@{}}$1-F^\infty$ \\ COBYLA\end{tabular} & \begin{tabular}[c]{@{}l@{}}$1-F^\infty$ \\ SPSA\end{tabular} & \begin{tabular}[c]{@{}l@{}}$1-F^\infty$ \\ Adam\end{tabular} & \begin{tabular}[c]{@{}l@{}}$1-F^\infty$ \\ L-BFGS-B\end{tabular} \\\hline
2 & RY1 & 2  & 1 & 0.0320(7) & 0.03182(7) & 0.03185(4) & $3.19\cdot 10^{-2}$ \\
2 & RY2 & 3 & 1 & \textbf{0.0310(8)} & 0.03184(3)  & 0.03185(0) & $3.19\cdot 10^{-2}$\\
2 & ZGR & 1 & 1 & 0.0319(7) & 0.03186(8) & 0.03188(0) & $3.19\cdot 10^{-2}$ \\

3 & RY1 & 4 & 3 & 0.002(3) & 0.00008(0) & 0.000060(8) & $5.89\cdot 10^{-5}$ \\
3 & RY2 & 6 & 4 & 0.001(4) & 0.000070(3)  & \textbf{0.000060(5)} & $5.89\cdot 10^{-5}$ \\
3 & ZGR & 3 & 4 & 0.0019(4)  & 0.0003(8) & 0.000069(9) & $5.89\cdot 10^{-5}$ \\

4 & RY1 & 6 & 6 & 0.005(8) & 0.0000(6) & 0.0000(2) & $2.13\cdot 10^{-5}$ \\
4 & RY2 & 9 & 9 & 0.006(8) & 0.0002(2)  & 0.0000(8) &  $1.72\cdot 10^{-10}$ \\
4 & ZGR & 7 & 9 & 0.006(6) & 0.000(1) & \textbf{0.000019(1)} & $1.73\cdot 10^{-9}$ \\

5 & RY1 & 8 & 10 & 0.02(7) & 0.0019(2) & 0.001(7) & $1.47\cdot 10^{-3}$ \\
5 & RY2 & 12 & 16 & 0.02(5) & 0.001(8)  & 0.003(4) & $4.51\cdot 10^{-8}$ \\
5 & ZGR & 15 & 18 & 0.01(2) & 0.002(6) & \textbf{0.000013(5)} & $6.23\cdot 10^{-8}$ \\

6 & RY1 & 10 & 15 & 0.09(3) & 0.007(4) & 0.006(2) & $5.71\cdot 10^{-3}$ \\
6 & RY2 & 15 & 25 & 0.14(1)  & 0.003(8) & 0.008(2) & $1.47\cdot 10^{-5}$ \\
6 & ZGR & 31 & 35 & 0.10(1) & 0.00(1) & \textbf{0.00006(9)} & $4.68\cdot 10^{-8}$ \\\hline\hline
\end{tabular}}}}
\caption{Noiseless results of the simulations for the harmonic oscillator for each number of qubits, \textit{Ansätze} and optimizers. The number of parameters and the number of CNOT gates of each \textit{Ansatz} for each number of qubits are shown. The infidelities for the COBYLA, SPSA and Adam optimizers are the median of the infidelities over 100 simulations using QISKIT QASM simulator with 8192 evaluations. The L-BFGS-B optimizer is combined with the statevector simulator to establish the numerical limit of each \textit{Ansatz}. We highlight in bold letter font the best result for each number of qubits.}
\label{tab:ho}
\end{table}

\begin{table}[H]
\tiny
\centering
{\setlength\arrayrulewidth{0.2pt}
{\fontsize{2}{3}\selectfont\resizebox{\textwidth}{!}{\begin{tabular}{c c c c c c c c}\hline \hline
Qubits & \textit{Ansätze} & Parameters &  CNOT gates & \begin{tabular}[c]{@{}l@{}}$1-F^\infty$ \\ COBYLA\end{tabular} & \begin{tabular}[c]{@{}l@{}}$1-F^\infty$ \\ SPSA\end{tabular} & \begin{tabular}[c]{@{}l@{}}$1-F^\infty$ \\ Adam\end{tabular} & \begin{tabular}[c]{@{}l@{}}$1-F^\infty$ \\ L-BFGS-B\end{tabular} \\\hline 
2 & RY1 & 2 & 1 & 0.159(5) & 0.15907(4) & 0.15910(0) & $1.59\cdot 10^{-1}$ \\
2 & RY2 & 3 & 1 & 0.1602(3) & 0.15909(8) & 0.15911(0) & $1.59\cdot 10^{-1}$ \\
2 & ZGR & 1 & 1 & 0.159(2) & \textbf{0.15906(5)} & 0.15917(8) & $1.59\cdot 10^{-1}$ \\
3 & RY1 & 4 & 3 & 0.003(1) & 0.00129(0) & 0.00128(4) & $1.28\cdot 10^{-3}$ \\
3 & RY2 & 6 & 4 & 0.003(1) & 0.00129(2)  & \textbf{0.001280(2)} & $1.28\cdot 10^{-3}$ \\
3 & ZGR & 3 & 4 & 0.0020(0) & 0.00134(4) & 0.001314(4) & $1.28\cdot 10^{-3}$ \\
4 & RY1 & 6 & 6 & 0.007(9) & 0.0001(8) & 0.0001(2) & $1.06\cdot 10^{-4}$\\
4 & RY2 & 9 & 9 & 0.010(1) & 0.0005(7)  & 0.0001(9) & $5.67\cdot 10^{-11}$ \\
4 & ZGR & 7 & 9 & 0.006(0) & 0.000(5) & \textbf{0.000109(6)} & $5.78\cdot 10^{-11}$ \\
5 & RY1 & 8 & 10 & 0.11(6) & 0.008(1) & 0.0066(3) & $6.15\cdot 10^{-3}$ \\
5 & RY2 & 12 & 16 & 0.12(5) & 0.002(7)  & 0.001(8) & $1.51\cdot 10^{-6}$ \\
5 & ZGR & 15 & 18 & 0.21(7) & 0.01(0) & \textbf{0.00022(7)} & $1.47\cdot 10^{-7}$\\
6 & RY1 & 10 & 15 & 0.271(1) & 0.01(9) & 0.0140(7) & $1.30\cdot 10^{-2}$ \\
6 & RY2 & 15 & 25 & 0.22(2)  & 0.01(9)  &  0.010(6) & $3.44\cdot 10^{-4}$ \\
6 & ZGR & 31 & 35 & 0.25(0) & 0.2(0) & \textbf{0.00031(4)} & $1.49\cdot 10^{-5}$ \\\hline\hline
\end{tabular}}}}
\caption{Noiseless results of the simulations for the transmon qubit for each number of qubits, \textit{Ansätze} and optimizers. We highlight in bold letter font the best result for each number of qubits.}
\label{tab:transmon}
\end{table}

\newpage

\begin{table}[H]
\centering
{\setlength\arrayrulewidth{0.2pt}
{\fontsize{2}{3}\selectfont\resizebox{\textwidth}{!}{\begin{tabular}{c c c c c c c c}\hline \hline
Qubits & \textit{Ansätze} & Parameters &  CNOT gates & \begin{tabular}[c]{@{}l@{}}$1-F^\infty$ \\ COBYLA\end{tabular} & \begin{tabular}[c]{@{}l@{}}$1-F^\infty$ \\ SPSA\end{tabular} & \begin{tabular}[c]{@{}l@{}}$1-F^\infty$ \\ Adam\end{tabular} & \begin{tabular}[c]{@{}l@{}}$1-F^\infty$ \\ L-BFGS-B\end{tabular} \\\hline
2 & RY1 & 2 & 1 & 0.0995(6) & 0.09919(4) & 0.09919(8) & $9.92\cdot 10^{-2}$ \\
2 & RY2 & 3 & 1 & 0.0996(0) & 0.09920(4) & 0.09919(6) & $9.92\cdot 10^{-2}$ \\
2 & ZGR & 1 & 1 & 0.0994(7) & \textbf{0.09918(4)} & 0.09921(8) & $9.92\cdot 10^{-2}$ \\
3 & RY1 & 4 & 3 & 0.04(7) & 0.06(6) & 0.0662(9) & $6.64\cdot 10^{-2}$ \\
3 & RY2 & 6 & 4 & \textbf{0.04(6)} & 0.06(5)  & 0.0662(2) & $6.64\cdot 10^{-2}$ \\
3 & ZGR & 3 & 4 & 0.056(1) & 0.057(9) & 0.0663(2) & $6.64\cdot 10^{-2}$ \\
4 & RY1 & 6 & 6 & 0.03(5) & 0.00(7) & 0.000(0) & $8.00\cdot 10^{-5}$\\
4 & RY2 & 9 & 9 & 0.06(0) & 0.00(2)  & 0.00(0) & $4.35\cdot 10^{-5}$ \\
4 & ZGR & 7 & 9 & 0.00(5) & 0.00(0) & \textbf{0.000051(4)} & $4.35\cdot 10^{-5}$ \\
5 & RY1 & 8 & 10 & 0.06(4) & 0.04(2) & 0.035(2) & $3.36\cdot 10^{-2}$ \\
5 & RY2 & 12 & 16 & 0.07(2) & 0.01(0)  & 0.00(0) & $6.65\cdot 10^{-5}$ \\
5 & ZGR & 15 & 18 & 0.02(4) & 0.04(0) & \textbf{0.000029(0)} & $6.96\cdot 10^{-8}$\\
6 & RY1 & 10 & 15 & 0.10(5) & 0.06(9) & 0.065(4) & $6.33\cdot 10^{-2}$ \\
6 & RY2 & 15 & 25 & 0.09(7)  & 0.04(7)  &  0.03(2) & $6.95\cdot 10^{-4}$ \\
6 & ZGR & 31 & 35 & 0.15(7) & 0.34(8) & \textbf{0.00010(3)} & $3.33\cdot 10^{-6}$ \\\hline\hline
\end{tabular}}}}
\caption{{Noiseless results of the simulations for the flux qubit for each number of qubits, \textit{Ansätze} and optimizers. We highlight in bold letter font the best result for each number of qubits.}}
\label{tab:flux qubit}
\end{table}

\twocolumngrid

\bibliography{my_bibliography}

\end{document}